\documentclass[structabstract,onecolumn]{aa}
\pdfoutput=1                        % (traditional abstract) 

\usepackage{graphicx}
\usepackage{amsmath,amssymb}
\usepackage{txfonts}
\usepackage{natbib} % uses number mode for citation
\usepackage{aas_macros} % Macros for journal names
\usepackage{url}
\bibpunct{(}{)}{;}{a}{}{,}

\newcommand{\bs}[1]{\boldsymbol #1}
\newcommand{\md}[0]{{\rm d}}

\newcommand{\rez}[1]{\frac{1}{#1}}

\newcommand{\av}[1]{\langle #1 \rangle}

\newcommand{\abl}[0]{\mathrm{d}}

\def\Mpc{\, h^{-1} \, {\rm Mpc}}

\def\H{\, {\rm km} \,\, {\rm s}^{-1} \, {\rm Mpc}^{-1}}
\def\Msun{\, h^{-1} \, M_\odot}

\begin{document}
\title{A fitting formula for the non-Gaussian contribution to the lensing
  power spectrum covariance}

%\subtitle{}

\author{J. Pielorz \and 
  J. R\"{o}diger \and
  I.Tereno \and 
  P. Schneider}

\institute{Argelander-Institut f\"ur Astronomie (AIfA), Universit\"at
  Bonn, Auf dem H\"ugel 71, 53121 Bonn, Germany\\
  \email{pielorz@astro.uni-bonn.de}
}

\date{Received; accepted}
 
\abstract
{Weak gravitational lensing is one of the most promising tools to
  investigate the equation-of-state of dark energy. In order to
  obtain reliable parameter estimations for current and future
  experiments, a good theoretical understanding of dark matter
  clustering is essential. Of particular interest is the statistical
  precision to which weak lensing observables, such as cosmic
  shear correlation functions, can be determined.}
{We construct a fitting formula for the non-Gaussian part of the
  covariance of the lensing power spectrum. The Gaussian contribution to the
  covariance, which is proportional to the lensing power spectrum
  squared, and optionally shape noise can be included easily by adding
  their contributions.}
{Starting from a canonical estimator for the dimensionless lensing power
  spectrum, we model first the covariance in the halo model approach including
  all four halo terms for one fiducial cosmology and then fit two polynomials
  to the expression found. On large scales, we use a first-order polynomial in
  the wave-numbers and dimensionless power spectra that goes asymptotically
  towards $1.1 \, {\cal C}_{\rm pt}$ for $\ell \rightarrow 0$, i.e., the
  result for the non-Gaussian part of the covariance using tree-level
  perturbation theory. On the other hand, for small scales we employ a
  second-order polynomial in the dimensionless power spectra for the fit.}
{We obtain a fitting formula for the non-Gaussian contribution of
  the convergence power spectrum covariance that is accurate to $10\%$ for
  the off-diagonal elements, and to $5\%$ for the diagonal
  elements, in the range $50 \lesssim \ell \lesssim 5000$ and can be
  used for single source redshifts $z_{\rm s} \in [0.5,2.0]$ in
  WMAP5-like cosmologies.}
{}
   

\titlerunning{A fitting formula for the lensing power spectrum covariance}
\authorrunning{J. Pielorz et al.}
\maketitle

\section{Introduction}
\label{sec:introduction}

Weak gravitational lensing by the large-scale structure, or cosmic shear,
is an important tool to probe the mass distribution in the
Universe and to estimate cosmological parameters. The constraints it
provides are independent and complementary to those found by other
cosmological probes such as cosmic microwave background (CMB)
anisotropies, supernovae (SN) type Ia, baryon acoustic oscillations
(BAO) or galaxy redshift surveys. The cosmic shear field quantifies the
distortion of faint galaxy images that is induced by continuous light
deflections caused by the large-scale structure in the Universe
\citep[e.g.,][]{Schneider2001,2006glsw.book..269S}. Since this effect
is too small to be measured for a single galaxy, large surveys with
millions of galaxies are required to detect it in a statistical way.
The cosmic shear signal has been successfully measured in various
surveys, since the first detections of 
\cite{2000MNRAS.318..625B,2000astro.ph..3338K,2000A&A...358...30V,2000Natur.405..143W}.
Most recently, shear two-point correlation functions were measured in
the Canada-France-Hawaii Telescope Legacy Survey (CFHTLS) 
and were used to constrain the amplitude of dark matter
clustering, $\sigma_8\Omega_{\rm m}^{0.6}$, with $5\%$ uncertainty
\citep{2008A&A...479....9F}.

The next generation of galaxy surveys will greatly improve the
precision with which weak lensing effects can be measured
\citep{2006astro.ph..9591A} enabling us to obtain, with accurate redshift
information and tomographic measurements, precise constraints on the
evolution of dark energy. However, the expected improvement in future
data only leads to a significant improvement of the precision and
accuracy of the cosmological interpretation if the systematic errors,
the underlying physics, and the statistical precision of cosmic shear
estimators are well understood. Systematics currently identified arise
mainly from non-cosmological sources of shear correlations, i.e.,
intrinsic alignments of galaxies \cite[e.g.,][for a
review]{2008arXiv0808.0203S}, and biases on the shear measurement
\citep{2007MNRAS.376...13M,2008arXiv0812.1881S}. This paper addresses
the issue of the statistical precision of cosmic shear estimators,
determined by the covariance of the estimator. Since much of the
scales probed by cosmic shear lie in the non-linear regime, being
affected by non-linear clustering, the covariance depends on
non-Gaussian effects and has a non-Gaussian, as well as a Gaussian,
contribution. Indeed, even though the non-Gaussianity of the shear
field is weaker than that of the matter field due to the projection along the line-of-sight,
various studies indicate that the non-Gaussian contribution to the
covariance cannot be neglected when constraining cosmological
parameters with weak lensing
\citep{1999ApJ...527....1S,2000ApJ...537....1W,CoorayHu2001,2005A&A...442...69K,2007MNRAS.375L...6S,2009MNRAS.395.2065T}.

Most cosmic shear results are based on the measurement of two-point
correlation functions of the shear field. Since, in general, the number of
independent measurements is insufficient to infer the complete covariance
directly from observations, one may derive it from ray-tracing maps of
numerical $N$-body simulations. This, however, requires a large number of
realizations and, in addition, is very time-consuming if an exploration of the
covariance in the parameter space is needed. An alternative is to compute the
covariance with an analytic approach. For shear two-point correlation
functions, \citet{2002A&A...396....1S} derived an expression for the Gaussian
contribution to the covariance. \citet{2007MNRAS.375L...6S} fitted the ratio
between that expression and a covariance computed with $N$-body simulations,
containing both Gaussian and non-Gaussian contributions, providing thus a
formula to compute the total covariance from the Gaussian term. In Fourier
space, large-scale modes are independent and, differently from real space, the
Gaussian contribution to the covariance of the convergence power spectrum
(i.e., of the Fourier transform of the two-point shear correlation function)
is diagonal and can be computed from the convergence power spectrum alone
\citep{1992ApJ...388..272K,2008A&A...477...43J}, whereas the non-Gaussian
contribution can be computed from the trispectrum of the convergence
\citep{1999ApJ...527....1S}. The trispectrum, on large scales, can be
accurately derived in tree-level perturbation theory\footnote{We refer by
  tree-level perturbation theory to the lowest, non-vanishing order of the
  considered quantity in perturbation theory.}, and, on small scales, is well
represented by the one-halo term of a halo model approach. A non-Gaussian part
of the covariance consisting of a perturbation theory term and a one-halo term
was used, e.g., in \citet{2009MNRAS.395.2065T}.

This paper aims at producing an accurate expression for the non-Gaussian
contribution of the covariance of the convergence power spectrum that is fast
to compute, contributing thus to accurate estimates of cosmological parameters. 
Following \citet{1999ApJ...527....1S} and \citet{CoorayHu2001}, we
start from a canonical estimator of the dimensionless convergence
power spectrum and use it to derive an analytic expression for the
corresponding covariance. The various spectra involved are evaluated using
the halo model approach of dark matter clustering
\citep{2000MNRAS.318..203S,2000ApJ...543..503M,2001ApJ...546...20S,
  2002PhR...372....1C}. 
The halo model approach assumes that all dark matter in the
Universe is bound in spherical halos, and uses results from
numerical $N$-body simulations to characterize halo properties such as
their profile, abundance and clustering behavior.

The evaluation of the covariance of the convergence power spectrum in
the halo model approach is time-consuming. In addition, it may be needed to
repeat it for different cosmological models for the purpose of
parameter estimation. To allow for a faster computation, we construct
a fitting formula for the non-Gaussian part of the convergence power
spectrum covariance. On small scales, we fit the halo model result
with a polynomial in the non-linear dimensionless convergence power
spectrum. On large scales, we fit the ratio between the halo model
covariance and the perturbation theory covariance. We stress that it
is a fit to the halo model covariance, not involving a covariance
computed from $N$-body simulations. The result is, however, calibrated
by $N$-body simulations, since they determine the halo model
parameters.

The paper is organized as follows. We define in
Sect.~\ref{sec:lambda-cdm-cosmology} the reference cosmology, considering the
growth of matter perturbations. We introduce the convergence spectra,
construct an estimator for the dimensionless convergence power spectrum, and
derive an expression for its covariance in Sect.~\ref{sec:con-power-spec-cov}.
In Sect.~\ref{sec:halo-model}, we describe the halo model approach, and
compute the covariance of the power spectrum. The covariance depends on the
values of halo model parameters, which are also defined here. It also depends
on the power spectrum, bispectrum and trispectrum of the correlations of halo
centers. Expressions for these spectra, in the framework of perturbation
theory, are given in the Appendix. Section~\ref{sec:comparison-with-n-body}
tests the accuracy of the halo model predictions, for both the power spectrum
and its covariance, against two sets of ray-tracing simulations.
Section~\ref{sec:fitting-formula} presents the fitting formula for the
non-Gaussian contribution to the covariance where its coefficients are given
as function of source redshift. We conclude in Sect.~\ref{sec:discussion}.

\section{Structure formation in a $\Lambda$CDM cosmology}
\label{sec:lambda-cdm-cosmology}

Throughout this work we assume a spatially flat cold dark matter model with a
cosmological constant ($\Omega_{\rm{m}}+\Omega_{\Lambda}=1$), as supported by
the latest 5-year data release of WMAP results \citep{WMAP5}. The expansion
rate of the Universe, $H(a) \equiv \dot{a}/a$, in such models is described by
the Friedmann equation $H^2(a)=H_0^2 \left(\Omega_{\rm m} a^{-3} +
  \Omega_\Lambda \right)$, where $H_0 \equiv 100 \, h \H$ is the Hubble
constant, $\Omega_{\rm m}$ denotes the combined contributions from dark matter
and baryons today in terms of the critical density $\rho_{\rm crit}\equiv
3H_0^{2}/(8\pi G)$, and $\Omega_\Lambda$ is the density parameter of the
cosmological constant. The comoving distance to a source at $a$
is then
\begin{equation}
  \label{eq:comovong_dist}
  w(a)= \int_a^1 \frac{c \, \md a'}{{a'}^2 H(a')} \,,
\end{equation}
where the scale factor is related to the redshift via the relation
$1+z=1/a$ using the convention $a(t_0)=1$ today.

In structure formation, the central quantity is the Fourier transform of the
density contrast $\delta(\bs x,t) = [\rho(\bs x,t) -
\bar{\rho}(t)]/\bar{\rho}(t)$, which describes the relative deviation of the
local matter density $\rho(\bs x,t)$ to the comoving average density of the
Universe $\bar{\rho}(t)$ at time $t$. We suppress the time dependence of
$\delta$ in the following. In this way, the mean density contrast is by
definition zero, and we can describe matter perturbations in the early
Universe as zero-mean Gaussian random fields. In this case, the statistical
properties of the Fourier transformed density field,
\begin{equation}
  \label{eq:fourier_density}
  \tilde{\delta}(\bs k)=\int \md^3 x \, {\rm e}^{{\rm i} \bs k \cdot \bs
    x} \delta(\bs x)\,, 
\end{equation}
are completely characterized by the power spectrum
\begin{equation}
  \label{eq:power_spectrum}
\av{\tilde{\delta}(\bs k_1)\tilde{\delta}(\bs k_2)} \equiv (2\pi)^{3}
  \delta_{\rm{D}}(\bs k_1 + \bs k_2) P_\delta(k_1) \,,
\end{equation}
where $\av{\cdot}$ is the ensemble average and $\delta_{\rm{D}}$
denotes the Dirac delta distribution. Note that throughout this paper
the tilde symbol is used to denote the Fourier transform of the
corresponding quantity. 

In linear perturbation theory, which is valid
on large scales, the power spectrum at a scale factor $a$ is
characterized by
\begin{equation}
  \label{eq:lin_power}
  P_{\rm lin}(k,a) =A \, k^{n_{\rm{s}}} T^{2}(k) D^2(a) \,, 
\end{equation}
where the amplitude $A$ is normalized in terms of $\sigma_8$, $n_{\rm{s}}$
denotes the spectral index of the primordial power spectrum, and $T(k)$ is the
transfer function. Note that all Fourier modes of the matter density grow at
the same rate, i.e., $\tilde{\delta}(\bs k,a)=\tilde{\delta}(\bs k)D(a)$,
where
\begin{equation}
  \label{eq:growth_factor}
  D(a) \propto \frac{H(a)}{H_0} \int_0^a
  \frac{\md a'}{[a'H(a')/H_0]^3} \,
\end{equation}
is the growth factor which we normalize as $D(a=1)=1$. In the non-linear
regime, i.e., on small scales, different Fourier modes couple and the Gaussian
assumption cannot be maintained. Thus we have to consider higher-order moments
of the density field to describe its statistical properties. In perturbation
theory, it is possible to find analytic expressions for these moments, which
hold up to the quasi-linear regime. In Appendix \ref{sec:PT}, we derive the
expressions for the bispectrum and trispectrum in tree-level perturbation
theory, which are the Fourier transforms of the three- and
four-point-correlation functions, respectively.

\section{Covariance of the convergence power spectrum}
\label{sec:con-power-spec-cov}

A central quantity in weak lensing applications is the two-dimensional
projection of the density contrast $\delta(w \bs \theta, w)$ on the sky,
which is known as \emph{effective convergence} $\kappa(\bs \theta)$.
It is obtained by projecting the density contrast along the 
 backward-directed light-cone of the observer according to
\begin{equation}
  \label{eq:eff-convergence}
  \kappa(\bs \theta)=\int_0^{w_{\rm H}} \md w \, w \, G(w) \, 
  \delta(w \bs \theta, w) \,, 
\end{equation}
where $w \equiv w(z)$ denotes the redshift-dependent comoving
distance, $w_{\rm H}$ is the comoving distance to the horizon and the
weight function $G(w)$ takes into account the distribution of source
galaxies along the line-of-sight. We assume for simplicity that all
background sources are situated at a single comoving distance $w_{\rm
  s} \equiv w(z_{\rm s})$, such that the weight function has the form
\begin{equation}
  \label{eq:weight_function}
  G(w)= \frac{3}{2}  \, \Omega_{\rm m} \, \left(\frac{H_0}{c}\right)^{2}a^{-1} \,
  \frac{w_{\rm s}-w}{w_{\rm s}} \operatorname{H}(w_{\rm{s}}-w)\,, 
\end{equation}
where $\operatorname{H}(x)$ denotes the Heaviside step function. To take
advantage of the Fourier properties, we analyze the statistical properties of
the Fourier counterpart of $\kappa(\bs \theta)$. For the theoretical
consideration of the convergence power spectrum covariance we need the second-
and the fourth-order moments, as will become apparent later. They are defined
by
\begin{align}
\label{eq:second_moment}
  \av{\tilde{\kappa}(\bs \ell_1)\tilde{\kappa}(\bs \ell_2)} &\equiv (2\pi)^{2}
  \delta_{\rm{D}}(\bs \ell_{12}) P_\kappa(\bs \ell_1) \,, \\
  \label{eq:fourth_moment}
  \av{\tilde{\kappa}(\bs \ell_1)\tilde{\kappa}(\bs \ell_2)\tilde{\kappa}(\bs \ell_3)\tilde{\kappa}(\bs
    \ell_4)}_{\rm c} &\equiv (2\pi)^{2} \delta_{\rm{D}}(\bs
    \ell_{1234}) T_\kappa(\bs \ell_1,\bs \ell_2, \bs \ell_3, \bs \ell_4)\,,
\end{align}
where the subscript `c' refers to the connected part of the corresponding
moment and $\bs \ell_{i \dots j}= \bs \ell_i + \ldots + \bs \ell_j$ is a sum
of Fourier wave-vectors. The convergence power spectrum and trispectrum are
calculated using the flat-sky and Limber's approximation
\citep{1998ApJ...498...26K,1999ApJ...527....1S,LSS_PT}:
\begin{align}
  \label{eq:pkappa}
  P_\kappa(\ell) &= \int_0^{w_{\rm H}}  \md w \, G^2(w) P_\delta
  \left(\frac{\ell}{w},w\right) \,, \\
  \label{eq:trikappa}
  T_\kappa(\bs \ell_1, \bs \ell_2, \bs \ell_3, \bs \ell_4 ) &= \int_0^{w_{\rm H}}
  \md w \, \frac{ G^4(w)}{w^2} \, T_{\delta}\left(\frac{\bs \ell_1
  }{w},\frac{\bs \ell_2}{w},\frac{\bs \ell_3}{w},\frac{\bs
    \ell_4}{w},w\right) \,,
\end{align}
where $P_\delta$ and $T_\delta$ are the corresponding
three-dimensional matter power spectrum and trispectrum (Fourier transform of
the four-point correlation function).

We are interested in estimating the dimensionless convergence power
spectrum
\begin{equation}
{\cal P_\kappa}(\ell)=(\ell^2/2\pi) \, P_\kappa(\ell)\,,
\label{eq:dimpk}
\end{equation}
and the
corresponding covariance for wave-vectors of different length $\ell$.
A natural choice for the
estimator of the dimensionless convergence power spectrum is
(\citealt{1999ApJ...527....1S,CoorayHu2001,2007NJPh....9..446T})
\begin{equation}
  \label{eq:estimator}
  \hat{\cal P}_\kappa(\ell_{i})=\rez{A} \int_{|\bs \ell| \in \ell_i }
  \frac{\md^{2}\ell}{A_{\rm{r}}(\ell_{i})} \frac{\ell^2}{2\pi} \, \tilde{\kappa}(\bs \ell) \tilde{\kappa}(-\bs \ell) \,,
\end{equation}
which is unbiased in the limit of infinitesimal small bin sizes, since
$\av{\hat{\cal P}_\kappa(\ell_{i})} \equiv {\cal P_\kappa}(\ell_{i})$. Here,
$A=4\pi f_{\rm sky}$ denotes the solid angle of a survey with a fractional sky
coverage of $f_{\rm sky}$, and the integration is performed over the Fourier
modes lying in the annulus defined by $\ell_i - \Delta \ell_i/2 \leq \ell \leq
\ell_i + \Delta \ell_i/2$, where $\Delta \ell_{i}$ is the width of the $i$-th
bin. We denote the integration area formed by the annulus as
$A_{\rm{r}}(\ell_{i})$.

The evaluation of the covariance of the estimator, Eq.~(\ref{eq:estimator}),
results in an expression of the form
\begin{align}
  \mathcal{C}_{ij} &\equiv {\rm Cov} \left[\hat{\cal P}_\kappa(\ell_{i}),
    \hat{\cal P}_\kappa(\ell_{j}) \right] =\langle\hat{\cal P}_\kappa(\ell_{i}) \hat{\cal
    P}_\kappa(\ell_{j})\rangle -\langle \hat{\cal P}_\kappa(\ell_{i}) \rangle \langle
  \hat{\cal P}_\kappa(\ell_{j}) \rangle \notag \\
  &=\frac{1}{A}\left[ \frac{(2\pi)^{2}}{A_{\rm{r}}(\ell_{i})} 2 {\cal
      P}_\kappa^{2}(\ell_{i})\delta_{\ell_{i}\ell_{j}}+ \bar{T}_\kappa(\ell_{i},\ell_{j}) \right]
  \equiv \mathcal{C}^{\rm G}_{ij} + \mathcal{C}^{\rm NG}_{ij} \,,
\label{eq:wlcov}
\end{align}
where ${\cal P_\kappa}(\ell_i)$ is given by Eq.~(\ref{eq:dimpk}), and
$\delta_{\ell_{i}\ell_{j}}$ denotes the Kronecker delta. The second term in square brackets
is the bin-averaged convergence trispectrum,
\begin{equation}
\label{eq:tkappabar}
\bar{T}_\kappa(\ell_{i},\ell_{j}) = \int_{|\bs
  \ell_1 | \in \ell_i } \frac{\abl^{2}\ell_{1}}{A_{\rm{r}}(\ell_{i})} \int_{|\bs \ell_2| \in \ell_j } \frac{\abl^{2}\ell_{2}}{A_{\rm{r}}(\ell_{j})} \, \frac{\ell_1^2
  \ell_2^2}{(2\pi)^2} \,
{T}_\kappa (\bs \ell_{1}, -\bs \ell_{1}, \bs \ell_{2}, -\bs \ell_{2}) \,,
\end{equation}
where $T_\kappa$ is the non-linear convergence trispectrum as defined in
Eq.~(\ref{eq:trikappa}). To derive Eq.~(\ref{eq:wlcov}), we made use of the
definitions of the convergence power spectrum and trispectrum in
Eqs.~(\ref{eq:second_moment}) and \eqref{eq:fourth_moment}, and of the
discrete limit of the delta distribution $\delta_{\rm{D}}(\bs 0) \rightarrow
A/(2\pi)^{2}$. The derived expression consists of two terms: a Gaussian part
$\mathcal{C}^{\rm G}$, which scales as the convergence power spectrum squared
and only contributes to the diagonal of the covariance matrix \citep[see,
e.g.][]{2008A&A...477...43J}, and a non-Gaussian part $\mathcal{C}^{\rm NG}$,
which scales as the dimensionless bin-averaged convergence trispectrum and
introduces correlations between the wave-vectors of different bins \citep[see,
e.g.][]{1999ApJ...527....1S,CoorayHu2001}. Both terms are inversely
proportional to the survey area $A$, but have a different behavior with
respect to the bin width $\Delta \ell_{i}$. While the Gaussian term
decreases with increasing bin size, the non-Gaussian term is independent of
the binning, since the bin area cancels out after the integration.

We can analytically perform one of the integrations of the bin-averaged
trispectrum in Eq.~\eqref{eq:tkappabar}. First, note that it only depends on
the parallelogram configuration of the convergence trispectrum, i.e., setting
$\bs \ell_2=-\bs \ell_1$, $\bs \ell_3=\bs \ell_2$ and $\bs \ell_4=-\bs \ell_2$
in Eq.~\eqref{eq:trikappa}. Also, if we choose an appropriate coordinate
system for the integration over the wave-vectors, the problem becomes
symmetric under rotations and we can parametrize the convergence trispectrum
by the length of the two sides of the parallelogram $\ell_1$ and $\ell_2$ and
the angle between them, $\cos \varphi = (\bs \ell_1 \cdot \bs \ell_2)/\ell_1
\ell_2$. Hence, we define
\begin{equation}
  \label{eq:lensing_tri_para}
  T_\kappa(\ell_1,\ell_2,\cos \varphi) \equiv T_\kappa (\bs \ell_{1}, -\bs
  \ell_{1}, \bs \ell_{2}, -\bs \ell_{2}) \,.
\end{equation}
Making use of the symmetry properties of this problem, one
angular integration becomes trivial and the integration in
Eq.~(\ref{eq:tkappabar}) simplifies to
\begin{equation}
  \label{eq:tkappabar_simp}
  \bar{T}_\kappa(\ell_{i},\ell_{j}) = \rez{2\pi}
  \int_{|\bs\ell_1| \in \ell_i} \frac{\md \ell_{1}}{A_{\rm{r}}(\ell_i)} \, \ell_1^3  
  \int_{|\bs\ell_2| \in \ell_j} \frac{\md \ell_{2}}{A_{\rm{r}}(\ell_j)} \, \ell_2^3  
  \int_{0}^{2\pi} \md \varphi \, {T}_{\kappa} (\ell_{1},\ell_{2},\cos\varphi) \,.
\end{equation}
If the bin-width $\Delta \ell_{i}$ is sufficiently small ($\Delta\ell_i \ll
\ell_i$), the integration area is $A_{\rm{r}}(\ell_{i})=2\pi \ell_{i} \Delta
\ell_i $, and we can make use of the mean value theorem. In this way, we
approximate the integral in Eq.~(\ref{eq:tkappabar_simp}) by an angular
average
\begin{equation}
  \label{eq:approx}
  \bar{T}_\kappa(\ell_{i},\ell_{j}) \simeq  \frac{1}{2\pi}\int_{0}^{2\pi} \md \varphi \,
  \frac{\ell_i^2 \ell_j^2}{(2\pi)^2} {T}_{\kappa}(\ell_{i},\ell_{j},\cos \varphi) \,.
\end{equation}
Note that if ${T}_{\kappa}(\ell_{i},\ell_{j},\cos \varphi)$ is independent of
the angle between $\bs \ell_i$ and $\bs \ell_j$, an approximation of the
covariance can be calculated without having to perform an integration at all.
In particular, this is the case for the 1-halo term of the three-dimensional
matter trispectrum, as we will see later in Eq.~(\ref{eq:tri_pc_1h}).

\section{Halo Model}
\label{sec:halo-model}

We have seen that the covariance of the dimensionless convergence
power spectrum estimator consists of two terms: a Gaussian part, which
is proportional to the dimensionless convergence power spectrum
squared and a non-Gaussian part, which is the bin-averaged
dimensionless convergence trispectrum (see Eq.~\ref{eq:wlcov}).
We will compute these terms using the halo model
approach
(\citealt{2000MNRAS.318..203S,2000ApJ...543..503M,2001ApJ...546...20S};
see also the comprehensive review by \citealt{2002PhR...372....1C}).

\subsection{Overview}
\label{sec:motivation}
 
With the assumption that all dark matter is bound in spherically-symmetric,
virialized halos, the halo model provides a way to calculate the
 three-dimensional
polyspectra of dark matter in the non-linear regime. In
Sect.~\ref{sec:trispectrum-2} below, we summarize the equations one
obtains for the dark matter power spectrum and trispectrum.

 In the halo model
description, the density field at an arbitrary position $\bs x$ in
space is given as a superposition of all $N$ halo density profiles such that
\begin{equation}
  \label{eq:superposition_halos}
  \rho(\bs x) = \sum_{i=1}^N  f(\bs x - \bs x_i;m_i,c_i) \equiv \sum_{i=1}^N m_i
  \, u(\bs x - \bs x_i;m_i,c_i) \,, 
\end{equation}
where $f(\bs x - \bs x_i;m_i,c_i)$ denotes the density profile of the $i$-th
halo with center of mass at $\bs x_i$ and $u \equiv f/m_i$ is the normalized
profile. By parametrizing the halo profile in this way, we assume that the
shape of the $i$-th halo depends only on the halo mass $m_i$ and the halo
concentration parameter $c_i$, which we define below. The dark matter
polyspectra of the density field $\rho(\bs x)$ follow then from taking the
ensemble averages, $\av{X}$, of products of the density at different points in
space. Assuming that the number of halos is $N=\bar{n}V$, where $\bar{n}$ is
the average number density of halos and $V$ the considered volume, we compute
the ensemble averages by integrating over the joint probability density
function (PDF) for the $N$ halos that form the field
\citep{2005MNRAS.360..203S}, i.e.,
\begin{align}
  \label{eq:ensemble_average}
  \av{X} \equiv \int \left[ \prod_{i=1}^{N} \md^3 x_i \, \md m_i \,
  \md c_i \right] 
   p(\bs x_1,\dots,\bs x_N, m_1,\dots,m_N,c_1,\dots,c_N) \, X \,,     
\end{align}
where $p(\cdot,\cdot,\cdot)$ denotes the PDF. If one considers that position
and mass of a single halo are independent random variables, the PDF factorizes as
\begin{equation}
  \label{eq:single_halo}
  p(\bs x;m,c)=p(\bs
  x)p(m)p(c|m)=\rez{V}\frac{n(m)}{\bar{n}}p(c|m) \,,
\end{equation}
where $n(m)$ is the halo mass function and $p(c|m)$ is the concentration
probability distribution for halos given a mass $m$. 

\subsection{Ingredients}
\label{sec:ingredients}

The halo model approach provides a scale-dependent description of the
statistical properties of the large-scale structure. On small scales, the
correlation of dark matter is governed by the mass profiles of the halos,
whereas on large scales the clustering between different halos determines the
nature of the correlation. As there are a multitude of models to describe the
behavior on different scales, and an even larger number of parameters one has
to set judiciously, there exists no such thing as a unique halo model. In
order to have reproducible results, it is therefore necessary to specify ones
choice of parameters. For this work, we will adopt the following parameters
for the halo model:
\begin{enumerate}
\item The average mass of a halo is defined as the mass within a
  sphere of virial radius $r_{\rm vir}$ as $m\equiv(4\pi/3)r_{\rm
    vir}^3 \, \Delta_{\rm vir} \, \bar\rho$, where $\Delta_{\rm vir}$
  denotes the overdensity of the virialized halo with respect to the
  average comoving mass density $\bar\rho$ in the Universe. Typically,
  values for $\Delta_{\rm vir}$ are derived in the framework of the
  non-linear spherical collapse model
  \citep[e.g.][]{1972ApJ...176....1G}. Expressions valid for different
  cosmologies are summarized in \citet{1997PThPh..97...49N}. In our
  implementation, we use the results which are valid for a flat
  $\Lambda$CDM-Universe, i.e.,
  \begin{equation}
    \label{eq:delta_vir}
    \Delta_{\rm vir}(z) = 18\pi^2(1+0.4093 x^{2.71572}) \,,
  \end{equation}
  where $x \equiv (\Omega_{\rm m}^{-1}-1)^{1/3}/(1+z)$. We find for our
  fiducial WMAP5-like cosmology $\Delta_{\rm{vir}}(z=0)=349$.
\item $N$-body simulations suggest that the density profile of a halo
  follows a universal function. We choose to use the NFW profile
  (\citealt{NFW}), which is
  in good agreement with numerical results and has an analytical
  Fourier transform. It is given by
  \begin{equation}
    \label{eq:halo_nfw}
    \rho(r,m) = \frac{\rho_{\rm s}}{(r/r_{\rm s})(1+r/r_{\rm s}
      )^2} \,,
  \end{equation}
  where $\rho_{\rm s}$ is the amplitude of the density profile and
  $r_{\rm s}$ characterizes the scale at which the slope of the
  density profile changes. For small scales ($r \lesssim r_{\rm
    s}$) the profile scales with $\rho \propto r^{-1}$,
  whereas for large scales it behaves as $\rho \propto r^{-3}$. The Fourier transform of the NFW profile is
  \begin{align}
  \label{eq:nfw_analytic}
  \tilde{u}(k;m,c) &= \int \md^3 x \, \rho(\bs x; m,c) \, {\rm e}^{\rm i
    \bs k \cdot \bs x} \Bigg/ \int \md^3 x \, \rho(\bs x; m,c) =
  \int_0^{r_{\rm vir}} \md r \, 4\pi r^2 \frac{\sin(kr)}{kr}
  \frac{\rho(r;m,c)}{m} \notag \\
  &= \left[\ln(1+c) - \frac{c}{1+c} \right]^{-1} \left\{ \sin\eta
  \left[ \mathrm{Si}([1+c] \eta) - \mathrm{Si}(\eta) \right] +
  \cos\eta \left[ \mathrm{Ci}([1+c] \eta) - \mathrm{Ci}(\eta)
  \right] - \frac{\sin(c\eta)}{(1+c) \eta} \right\} \,,
\end{align}
where $\eta \equiv k r_{\rm vir}/c$, we truncated the integration at
$r_{\rm vir}$ in the second step, and introduced the concentration
parameter $c\equiv r_{\rm vir}/r_{\rm s}$ in the third step.
Additionally, we use for the sine- and cosine integrals the definitions
\begin{equation}
  \label{eq:si_ci_def}
  \mathrm{Ci}(x) = - \int_x^\infty \frac{\cos t}{t} \, \md t \,,\qquad
      \mathrm{Si}(x) = \int_0^x \frac{\sin t}{t} \, \md t \,.
\end{equation}
 \item The abundance of halos of mass $m$ at a redshift $z$ is given by
   \begin{equation}
    \label{eq:mass_function}
    n(m,z)=\frac{\bar{\rho}}{m^{2}}\, 
    \frac{\abl \ln\nu}{\abl \ln m} \,\nu f(\nu)\,,
  \end{equation}
  where we introduced the dimensionless variable $\nu=\nu(m,z)$,
  \begin{equation}
    \nu(m,z) \equiv \frac{\delta_{\rm{sc}}(z)}{D(z)\sigma(m)}\,,
  \end{equation}
  where $D(z)$ denotes the redshift-dependent growth factor,
  $\sigma^2(m)$ is the smoothed variance of the density contrast, and
  $\delta_{\rm sc}(z)$ denotes the value of a spherical overdensity
  that collapses at a redshift $z$ as calculated from linear
  perturbation theory. In our work, we use the expression from
  \citet{1997PThPh..97...49N}, which is valid for a $\Lambda$CDM
  Universe,
  \begin{equation}
  \label{eq:delta_sc}
    \delta_{\rm sc}(z) = \frac{3(12\pi)^{2/3}}{20} \left[1-0.0123
    \ln(1+x^3)\right] \,,
  \end{equation}
  where $x \equiv (\Omega_{\rm m}^{-1}-1)^{1/3}/(1+z)$. The quantity
  has only a weak dependence on redshift and we find
  $\delta_{\rm{sc}}(z=0)=1.675$ for our fiducial model.

  The advantage of introducing $\nu$ is that part of the mass function can be
  expressed by the multiplicity function $\nu f(\nu)$, which has a universal
  shape, i.e., is independent of cosmological parameters and redshift. In this
  work, we employ the Sheth and Tormen mass function \citep{ST1999}
  \begin{equation}
    \label{eq:halo_st_massfunction}
    \nu f(\nu) = A(p) \left[1+(q\nu^2)^{-p} \right]
    \sqrt{\frac{2q}{\pi}} \, \nu \exp{(-q\nu^2/2)}\,,
  \end{equation}
  which is an improvement over the original Press-Schechter
  formulation \citep{PS}. We use the parameter values $p=0.3$, $q=0.707$,
  and amplitude $A(0.3)=0.322$, which follows from mass
  conservation.
  
\item The concentration parameter $c\equiv r_{\rm vir}/r_{\rm s}$
  characterizes the form of the halo profile. From $N$-body
  simulations one finds that the average, $\bar{c}$, depends on the
  halo mass \citep{Bullock2001} like
  \begin{equation}
    \label{eq:halo_conc}
    \bar{c}(m,z) = \frac{c_*}{1+z} \left(\frac{m}{m_*}\right)^{-\alpha} \,, 
  \end{equation}
  where $m_*=m_*(z=0)$ is the characteristic mass defined within the
  Press-Schechter formalism as $\delta_{\rm sc}(z=0)= \sigma(m_*)$. In
  the following, we will use the values $c_*=10$ and $\alpha=0.2$ as proposed
  by \citet*{TJ3PCF}. This implies that more massive halos are less centrally
  concentrated than less massive ones. 
  However, results from numerical $N$-body simulations
  \citep{2000ApJ...535...30J,Bullock2001} indicate that there is a significant
  scatter in the concentration parameter for halos of the same mass.
  Furthermore, \citet*{2000ApJ...535...30J} proposes that such a concentration
  distribution can be described by a log-normal distribution
  \begin{equation}
    \label{eq:conc_dist}
    p(c|m) \md c = \rez{\sqrt{2\pi\sigma^2_{\ln c}}} \exp \left[- \frac{(\ln c -
        \ln \bar c)^2}{2\sigma^2_{\ln c}}\right] \md \ln c \,.
  \end{equation}
  Typical values for the concentration dispersion range from $\sigma_{\ln
    c}=0.18$ to $\sigma_{\ln c}=0.32$
  \citep{2000ApJ...535...30J,2002ApJ...568...52W}. Note that the width of the
  distribution $\sigma_{\ln c}$ is independent of the halo mass. The variation
  of the halo concentration can be attributed to the different merger
  histories of the halos (\citealt{2002ApJ...568...52W}). We will analyze the
  impact of this effect on different spectra in
  Sect.~\ref{sec:stoch-halo-conc}. When we use only the mean concentration
  parameter, we have to replace the probability distribution of the
  concentration, needed for example in Eq.~\eqref{eq:single_halo}, by a Dirac
  delta distribution
  \begin{equation}
    \label{eq:conc_dirac}
    p(c|m) \md c = \delta_{\rm{D}}(c-\bar{c}) \,c \, \md \ln c\,.
  \end{equation}

\item On large scales, the correlation of the dark matter density
  field is governed by the spatial distribution of halos. Since the
  clustering behavior of halos and matter density differ, one
  introduces the bias factors $b_i(m,z)$ such that
    \begin{equation}
      \label{eq:bias_func}
      \delta_{\rm h}(\delta) \equiv \delta_{\rm h}(\bs x;m,z) =
      b_1(m,z) \delta(\bs x) + \frac{b_2(m,z)}{2} \delta^2(\bs x) +
      \dots 
    \end{equation}
    In this way, the halo density contrast, $\delta_{\rm h}(\delta)$, is
    expressed as a Taylor expansion of the matter density contrast,
    $\delta(\bs x)$. The bias parameters are in general derived based on the
    Sheth-Tormen mass function introduced above. For the linear halo bias one
    obtains then
    \begin{equation}
      \label{eq:halo_st_bias}
      b_1(m,z)= 1 + \frac{q[\nu(m,z)]^{2} -1}{\delta_{\rm sc}(z)} +
      \frac{2p}{\delta_{\rm sc}(z)[1+(q[\nu(m,z)]^{2})^p]} \,,
    \end{equation}
    where $p$ and $q$ match the values used in the mass function. Expressions
    for higher-order bias factors can be found, e.g., in
    \citet{2001ApJ...546...20S}. Since they only have a small impact on the
    quantities employed here, we take into account only the first-order bias.
    In Fourier space we may then write
    \begin{equation}
      \label{eq:bias_func_fourier}
      \tilde{\delta}_{\rm h}(\tilde{\delta}) \equiv
      \tilde{\delta}_{\rm h}(k;m,z) = b_1(m,z) \tilde{\delta}(k) \,.
    \end{equation}

  \item To obtain the final correlation function, one has to perform
    integrations along the halo mass and optionally along the halo
    concentration, with limits formally extending from $0$ to $\infty$. In
    practice, we use the mass limits $m_{\rm min}=10^3 \,h^{-1}\,M_\odot$ and
    $m_{\rm max}=10^{16} h^{-1}\,M_\odot$. Masses smaller than $m_{\rm
      min}=10^3 \,h^{-1}\,M_\odot$ give no significant contribution to the
    considered quantities, while, due to the exponential cut-off in mass,
    masses larger than $m_{\rm max}=10^{16} h^{-1}\,M_\odot$ are rare. For the
    concentration, we employ the integration limits $c_{\rm min}=1$ and
    $c_{\rm max}=10^3$.
 
\item Due to the cut-off in mass, the consistency relation \citep{2001ApJ...546...20S}
  \begin{equation}
    \label{eq:halo_consistency}
    \frac{1}{\bar{\rho}}\int_{m_{\rm min}}^{m_{\rm max}} \md m \, m \,
    n(m,z) \, b_{1}(m,z) = 1   
  \end{equation}
  does not hold. To cure this problem we consider a rescaled linear
  bias such that $b_{1}(m,z) \rightarrow b_{1}(m,z)/b_{\rm norm}(z)$,
  where $b_{\rm norm}(z)$ is the result of the integral in
  Eq.~(\ref{eq:halo_consistency}). In this way, one ensures that the
  halo term with the largest contribution to the correlation equals
  the perturbation theory expression on large scales (see
  Fig.~\ref{fig:trikappa}).

\end{enumerate}

\subsection{Building Blocks}
\label{sec:building-blocks}

Using the ingredients described in the previous section, it is possible to
define building blocks, which simplify significantly the notation for expressing
the polyspectra \citep{CoorayHu2001}:
\begin{align}
  \label{eq:mij_building_block}
  M_{ij} (k_1,\ldots,k_j;z) \equiv \int_{m_{\rm min}}^{m_{\rm max}}
  \md m \, \int_{c_{\rm min}}^{c_{\rm max}} \md c \, n(m,z) \, p(c|m)
  \left(\frac{m}{\bar\rho}\right)^j b_i(m,z) \, [\tilde{u}(k_1;m,c)
  \cdots \tilde{u}(k_j;m,c)] \,.
\end{align}
 In the case $i=0$, we additionally define $b_0 \equiv 1$, for consistency.

\subsection{Power spectrum}
\label{sec:power-spectrum}

We can now compute the power spectrum from Eq.~(\ref{eq:ensemble_average}).
The result consists of two terms, the 1-halo and the 2-halo terms,
$P_\delta(k)=P_{\rm 1h}(k)+P_{\rm 2h}(k)$ \citep{2000MNRAS.318..203S}. They are given by
\begin{align}
  \label{eq:power_spectrum_1h}
  P_{\rm 1h}(k) &= \rez{\bar\rho^2} \int_{m_{\rm min}}^{m_{\rm max}}
  \md m \,n(m) m^2 \int_{c_{\rm min}}^{c_{\rm max}} \md c \,  p(c|m) \,
   \, |\tilde{u}(k;m,c)|^2 \,, \\
  P_{\rm 2h}(k) &= \left[ \rez{\bar\rho} \int_{m_{\rm min}}^{m_{\rm
        max}} \md m \,n(m) m \,b_1(m)\int_{c_{\rm min}}^{c_{\rm max}} \md c \, 
    \, p(c|m)   \, \tilde{u}(k;m,c) \, \right]^2 P_{\rm
    lin}(k) \,,
\end{align}
where $P_{\rm lin}(k)$ denotes the linear perturbation theory power
spectrum, defined in Eq.~(\ref{eq:lin_power}), and we use the
ingredients summarized earlier-on. Note that, for convenience,
we omit the redshift-dependence in the notation.
Using the building blocks from
Eq.~(\ref{eq:mij_building_block}), these terms can be written in the following
compact form~:
\begin{align}
  P_{\rm 1h}(k) = M_{02}(k,k) \,, \qquad
  P_{\rm 2h}(k) = [M_{11}(k)]^2 P_{\rm lin}(k) \,.
\end{align}
The 1-halo term, $P_{\rm 1h}$, denotes correlations in space between two
points in the same halo, whereas the 2-halo term, $P_{\rm 2h}$, takes into
account correlations between two different halos.
Hence, the 1-halo term is dominant on small scales and the 2-halo term
is dominant on large scales. Note that the 2-halo term converges to
the linear power spectrum on large scales because of the consistency
relation of the first-order halo bias factor (see
Eq.~\ref{eq:halo_consistency}) and of the limit $\tilde{u}(k;m,c) \rightarrow 1$
for $k \rightarrow 0$.

\subsection{Trispectrum}
\label{sec:trispectrum-2}

We compute now the dark matter trispectrum in the halo model approach.
As discussed in Sect.~\ref{sec:con-power-spec-cov}, only parallelogram
configurations of the trispectrum wave-vectors contribute to the
covariance of the convergence power spectrum. Restricting our
calculations to these configurations, we obtain four different halo
term contributions,
\begin{equation}
  \label{eq:trispectrum_hm_pc}
T_{\delta}(k_1,k_2,\cos \varphi) = T_{\rm 1h}+T_{\rm 2h}+T_{\rm 3h}+T_{\rm 4h} \,,   
\end{equation}
which are simpler than in the general case. In addition, we neglect terms
involving higher-order halo bias factors since, on most scales, they provide
only a small correction \citep{2000ApJ...543..503M,TJ3PCF}. We further note
that the perturbative expansion of halo centers, used in the calculations, was
shown to become inaccurate on non-linear scales \citep{2007PhRvD..75f3512S}.
The contributions to the trispectrum take the following forms, using the
compact notation of the building blocks (see \citealt{2002PhR...372....1C} for
the expression of the halo model trispectrum including higher-order bias
factors): The 1-halo term, dominant on the smallest scales, is
\begin{equation}
\label{eq:tri_pc_1h}
  T_{\rm 1h} = M_{04}(k_1,-k_1,k_2,-k_2)  \,.
\end{equation}  
The 2-halo term has two contributions, $T_{\rm 2h} = T_{\rm 2h}^{31} + T_{\rm
  2h}^{22}$, consisting of a term $T_{\rm 2h}^{31}$, which corresponds
 to correlations of three points within one halo and a fourth point in a
 second halo, and a term
$T_{\rm 2h}^{22}$, which describes correlations involving two points in a
first halo and the other two points in the second halo. They read, 
\begin{align}
  T_{\rm 2h}^{31} &= 2 M_{13}(k_1,k_2,k_2) M_{11}(k_1) P_{\rm
    lin}(k_1)  + 2 M_{13}(k_1,k_1,k_2) M_{11}(k_2) P_{\rm lin}(k_2) \,,\\
  T_{\rm 2h}^{22} &= M_{12}^2(k_1,k_2) [P_{\rm lin}(|\bs k_1+ \bs k_2|)
  + P_{\rm lin}(|\bs k_1 - \bs k_2|)] \,.
\end{align}
The 3-halo term is given by
\begin{equation}
  \label{eq:tri_pc_3h}
 T_{\rm 3h} = 2 M_{12}(k_1,k_2) M_{11}(k_1) M_{11}(k_2) 
  \left[B_{\rm pt}(\bs k_1,\bs k_2,- \bs k_1 - \bs k_2)
    + B_{\rm pt}(\bs k_1,-\bs k_2,- \bs k_1 + \bs k_2) \right]\,.
\end{equation}
Finally, the 4-halo term, dominant on large scales, is
\begin{equation}
  \label{eq:tri_pc_4h}
  T_{\rm 4h} = M^2_{11}(k_1)
  M^2_{11}(k_2) T_{\rm pt}(\bs k_1,- \bs k_1,\bs k_2, -\bs k_2) \,,
\end{equation}
and describes correlations of points distributed in four different halos. Note
that, like for the power spectrum, the 2-halo term is computed from the linear
power spectrum. On the other hand, the 3- and 4-halo terms depend on $B_{\rm
  pt}$ and $T_{\rm pt}$, respectively, which are the lowest-order,
non-vanishing, perturbation theory contributions to the bispectrum and
trispectrum. Both spectra are derived in Appendix
\ref{sec:correlation-function}.

\subsection{Convergence spectra}

\begin{figure}
   \centering
   \includegraphics[angle=0,width=0.7\textwidth]{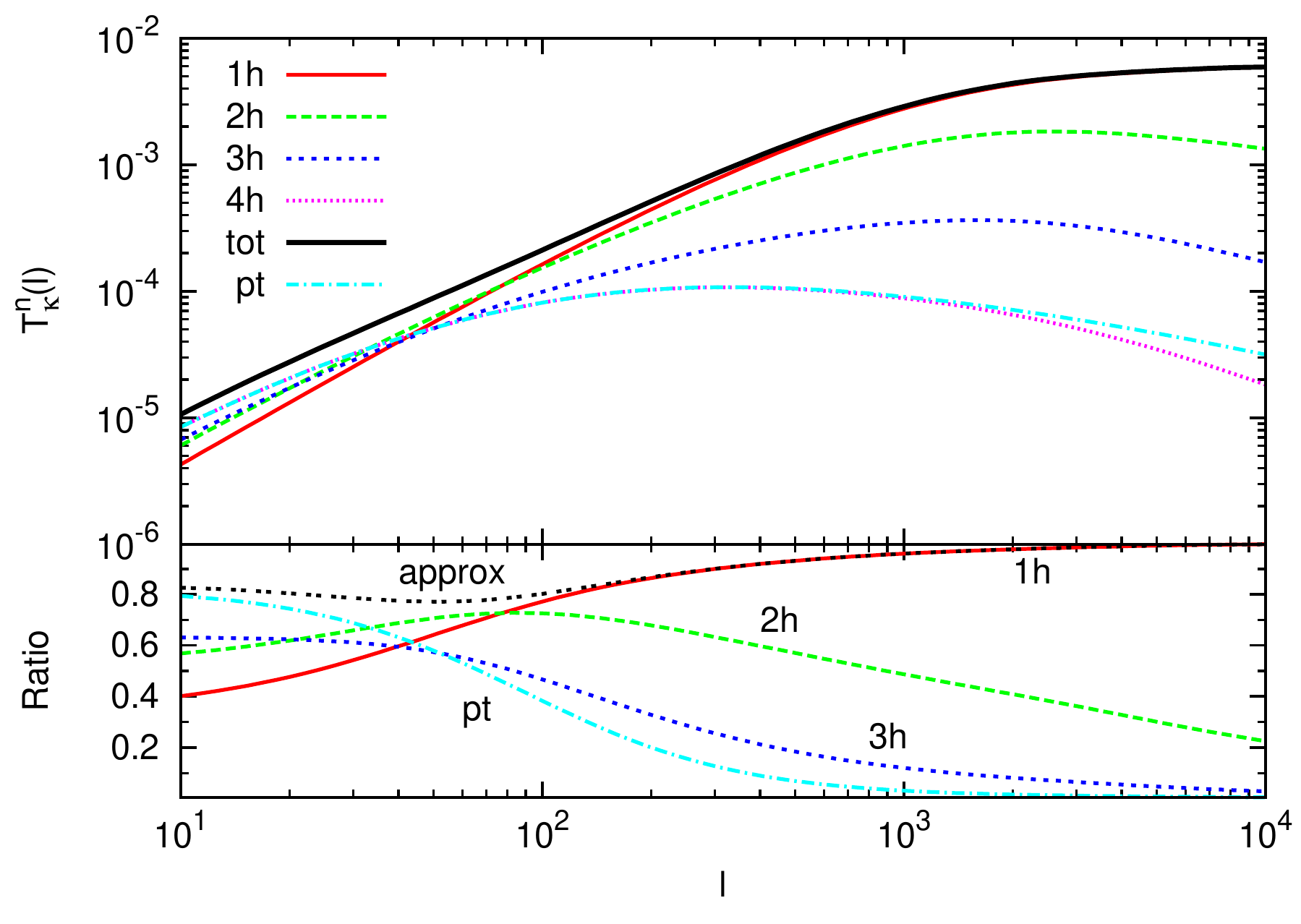}
   \caption{Square configuration of the dimensionless convergence
     trispectrum ${\cal T}^n_\kappa(\ell)$ against wave-number $\ell$
     for the range $10 \leq \ell \leq 10^4$ for $n \in \{\mbox{1h, 2h,
       3h, 4h, tot, pt}\}$. The upper panel displays the four
     individual halo terms, the sum of all four trispectrum halo terms
     ${\cal T}^{\rm tot}_\kappa$ (solid line) and the tree-level
     perturbation theory trispectrum ${\cal T}^{\rm pt}_\kappa$, as
     indicated in the key. The lower panel shows the ratio between the
     indicated contributions and the complete trispectrum. The double
     dashed line illustrates the corresponding ratio of the
     approximation ${\cal T}^{\rm tot}_\kappa \approx {\cal T}^{\rm
       pt}_\kappa + {\cal T}^{\rm 1h}_\kappa$. Note that we consider
     the 4-halo term only in the upper panel since it resembles the
     term ${\cal T}^{\rm pt}_\kappa$ on large scales.}
   \label{fig:trikappa}
\end{figure}

The convergence power spectrum and trispectrum, needed to evaluate the
covariance in Eq.~(\ref{eq:wlcov}), are computed by projecting
$P_{\delta}$ and $T_{\delta}$, according to Eqs.~(\ref{eq:pkappa}) and
(\ref{eq:trikappa}). Fig.~\ref{fig:trikappa} shows the dimensionless
convergence trispectrum, defined as ${\cal T}_\kappa^n(\ell)=
\ell^2/2\pi \, \sqrt[3]{T_\kappa^n(\bs \ell, -\bs \ell, \bs \ell, -\bs
  \ell)}$ for a square configuration where all wave-numbers have a
length $\ell$, where $n \in \{ \rm 1h,2h,3h,4h\}$ denotes the
contribution from the corresponding halo term. The plot shows the
individual contributions to the dimensionless ${\cal T}_\kappa^n$ (the
projections of Eqs.~\ref{eq:tri_pc_1h}-\ref{eq:tri_pc_4h}),
illustrating on which scales the individual terms are important, as
well as the total contribution of all halo terms ${\cal T}^{\rm
  tot}_\kappa$. We show, in addition, the dimensionless projected
${\cal T}^{\rm pt}_\kappa$, which closely follows the 4-halo term. We
see that the commonly used approximation ${\cal T}^{\rm tot}_\kappa
\approx {\cal T}^{\rm pt}_\kappa + {\cal T}^{\rm 1h}_\kappa$ is
accurate for large wave-numbers ($\ell \gtrsim 10^3$) but has a
deviation of about $20\%$ from the complete trispectrum ${\cal T}^{\rm
  tot}_\kappa $ for small wave-numbers ($\ell \lesssim 10^2$).

\subsection{Stochastic halo concentration}
\label{sec:stoch-halo-conc}

The previous results were computed using the \emph{deterministic} 
concentration-mass relation of Eq.~(\ref{eq:halo_conc}).
We now analyze the impact of scatter in the halo
concentration parameter $c$ on the covariance of the convergence power
spectrum, using the \emph{stochastic}
concentration relation given by the the log-normal concentration
distribution of Eq.~(\ref{eq:conc_dist}).

\citet{CoorayHu2001}, analyzed the effect of a stochastic concentration
on the three-dimensional power spectrum and trispectrum and found
that the behavior of the corresponding 1-halo terms were increasingly
sensitive to the width of the concentration distribution for smaller
wave-numbers $k$. Furthermore, the effect of a stochastic
concentration relation was more pronounced for the trispectrum than
for the power spectrum, since the tail of the concentration distribution
is weighted more strongly in higher-order statistics. 

Performing the
same analysis for the projected power spectrum and trispectrum, we
find a similar trend as in the three-dimensional case, but with a
smaller sensitivity to the concentration width of the distribution on
small scales. For $\sigma_{\ln c}=0.3$, we find, in the case of the 1-halo
term of the power spectrum, a deviation from a deterministic
concentration relation of about $1\%-2\%$ for wave-numbers larger than
$\ell \sim 10^3$, whereas, for the 1-halo term of the trispectrum, the 
deviation is of the order of $10\%-15\%$ in the same $\ell$-range. Thus,
when considering the covariance of the convergence power spectrum, one
should take into account the concentration dispersion in the 1-halo
term of the trispectrum but can safely neglect it for the power
spectrum. Additionally, we find that a stochastic concentration has
only a small impact on the 2-halo terms of the power spectrum and
trispectrum \citep{pielorz:2008}.

\begin{figure}
   \centering
   \includegraphics[angle=0,width=0.85\textwidth]{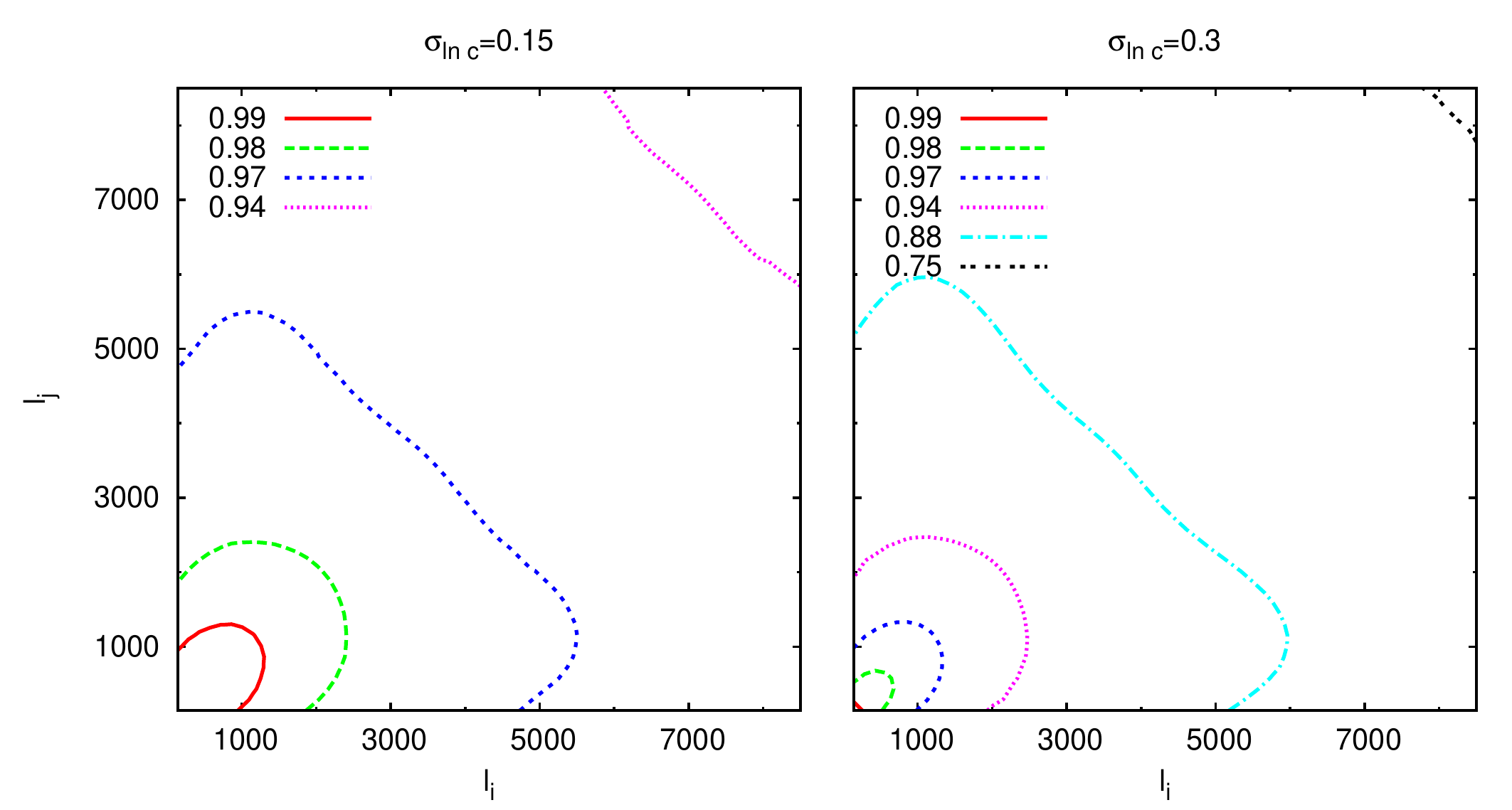}
   \caption{Contour plots of the ratio $R(\ell_i,\ell_j)$ between the
     non-Gaussian covariance 1-halo term contribution (see
     Eq.~\ref{eq:conc_ratio}) computed with a concentration dispersion
     $\sigma_{\ln c}$ and 
     with a deterministic concentration, against wave-numbers
     $(\ell_i,\ell_j)$ ranging from $\ell_i=\ell_j=500$ to
     $\ell_i=\ell_j=8500$. The left panel is for concentration
     dispersion $\sigma_{\ln c}=0.15$, whereas in the right panel
     $\sigma_{\ln c}=0.3$.}
   \label{fig:contours_varc}
\end{figure}

From this analysis, we expect the effect of a
concentration distribution to be the strongest on the non-Gaussian part of
 the covariance, which depends on the trispectrum. To directly infer the
impact of a concentration distribution on the covariance, we
calculate the 1-halo contribution to the non-Gaussian covariance, i.e., we
perform the bin averaging of Eq.~(\ref{eq:tkappabar}),
\begin{equation}
  \label{eq:cov_1h}
  \mathcal{C}_{1 \rm h}^{\rm{NG}}(\ell_i,\ell_j) \simeq
  \int_{{|\bs l_1|} \in \ell_i} \frac{\md \ell_{1}}{A_{\rm{r}}(\ell_i)} \, \ell_1^3  
  \int_{{|\bs l_2|} \in \ell_j} \frac{\md \ell_{2}}{A_{\rm{r}}(\ell_j)} \, \ell_2^3  
  \, {T}_{\kappa}^{1 \rm h}(\ell_{1},\ell_{2})\,,
\end{equation}
for two different concentration dispersions $\sigma_{\ln c}$, using our halo
model implementation.
Fig.~\ref{fig:contours_varc} shows the ratio

\begin{equation}
  \label{eq:conc_ratio}
  R(\ell_i,\ell_j) \equiv \frac{\mathcal{C}_{1 \rm h}^{\rm{NG}}(\ell_i,\ell_j;\sigma_{\ln
      c})}{\mathcal{C}_{1 \rm h}^{\rm{NG}}(\ell_i,\ell_j;\sigma_{\ln c}=0)} \,,
\end{equation}
where $\sigma_{\ln c}=0$ denotes the deterministic concentration
relation, for two values
$\sigma_{\ln c}=\{0.15,0.3\}$ in two contour plots. 
In agreement
with the previous results, the largest impact of the concentration
dispersion occurs for wave-numbers larger than $\ell_i \simeq \ell_j
\simeq 4000$. For $\sigma_{\ln c}=0.15$, the deviation
of the covariance $\mathcal{C}_{1 \rm h}^{\rm{NG}}$ from the original
deterministic concentration relation is small, of about $3\%-6\%$. The effect
becomes non-negligible for $\sigma_{\ln c}=0.3$, where we find a
deviation of $12\%$ (for $\ell_i \simeq \ell_j \gtrsim 4000$) to
$25\%$ (for $\ell_i \simeq \ell_j \gtrsim 8000$).

\section{Comparison with $N$-body simulations}
\label{sec:comparison-with-n-body}

In Sect.~\ref{sec:halo-model}, we computed the power spectrum and
trispectrum of the dark matter fluctuations. Projecting them according
to Eqs.~(\ref{eq:pkappa}) and (\ref{eq:trikappa}), and inserting the
result in Eqs.~(\ref{eq:dimpk}), (\ref{eq:wlcov}), and
(\ref{eq:tkappabar}), we obtained the covariance of the convergence
power spectrum estimator.

To test the accuracy of the halo model predictions for the
statistics of the dark matter density field, we compare the
dimensionless convergence power spectrum 
and the corresponding covariance, calculated in the
halo model approach, with results from two different ray-tracing
simulations.

\begin{table}
  \renewcommand{\arraystretch}{1.2}
  \caption{Cosmological parameters used to set up the initial power
    spectrum, which determines how the simulation particles are
    distributed initially. The simulations employ either the BBKS \citep{BBKS}
    or the EH 
    \citep{1998ApJ...496..605E} transfer function. Convergence maps
    with source galaxies situated at $z_{\rm s}=1$ and $z_{\rm s}=2$
    were produced for both sets of simulations.}
\begin{center}
\begin{tabular}{c c c c c c c c c c}
  \hline \hline
  Simulation & $\Omega_{\rm m}$ & $\Omega_\Lambda$ & $h$ &
  $\Omega_{\rm b}$ & $\sigma_8$ &
  $n_{\rm{s}}$ & $\Gamma$ & $z_{\rm s}$ & $T(k)$ \\ \hline
  Virgo & 0.3 & 0.7 & 0.7 & 0.0 & 0.9 & 1.0 & 0.21 & 1 (2) & BBKS \\
  Gems & 0.25 & 0.75 & 0.7 & 0.04 & 0.78 & 1.0 & 0.14 & 1 (2) & EH \\ \hline 
\end{tabular}
\end{center}
\label{tab:cosmo_parameters}
\end{table}

\begin{table}
  \renewcommand{\arraystretch}{1.2}
  \caption{Parameters used for generating the two $N$-body simulations
    considered in this paper and for producing the resulting
    convergence maps with the multiple-lens-plane ray-tracing algorithm (see e.g.,
    \citealt{2000ApJ...530..547J,2008arXiv0809.5035H}). From left to right
    these are: the side length, $L_{\mbox{\scriptsize box}}$, of the
    cubic simulation box, the number of
    particles, $N_{\mbox{\scriptsize par}}$, used for the simulation, their mass, $m_{\mbox{\scriptsize par}}$, and the number of available
    convergence maps, $N_{\mbox{\scriptsize map}}$, with area
    $A_{\mbox{\scriptsize map}}$.}
  \begin{center}
    \begin{tabular}{c c c c c c c}
      \hline \hline
      Simulation  & $L_{\mbox{\scriptsize box}}/\Mpc$ &
      $N_{\mbox{\scriptsize par}}$ & $m_{\mbox{\scriptsize par}}/ \Msun$
      & $N_{\mbox{\scriptsize map}}$&  $A_{\mbox{\scriptsize
          map}}/(\mbox{deg})^2$ \\ \hline
      Virgo & 141.3 & $256^3$ & $1.4 \times 10^{10}$ & 200 & 0.25 \\
      Gems & 150.0 & $256^3$ & $1.4 \times 10^{10}$ & 220 & 16.00 \\ \hline
    \end{tabular}
  \end{center}
  \label{tab:simu_para}
\end{table}

\subsection{Virgo and Gems simulation}
\label{sec:virgo-gems-sim}

For our comparison, we chose one simulation from
\citet{Jenkins1998} and ten simulations from
\citet{2009arXiv0901.3269H}, which we denote in the following as Virgo
and Gems simulation, respectively. The Virgo simulation was carried
out in 1997 by the Virgo-Consortium for a $\Lambda$CDM cosmology (see
Tab.~\ref{tab:cosmo_parameters}) with $N_{\rm par}=256^3$ particles
in a periodic box of comoving side length $L_{\rm box}=141.3 \Mpc$
(see Tab.~\ref{tab:simu_para}). It uses the PP-/PM-code HYDRA, which
places subgrids of higher resolution in strongly clustered regions.
Structures on scales larger than $2 l_{\rm soft} \approx 40\, h^{-1}\,
\mathrm{kpc}$ can be considered as well resolved. The Gems simulations
were set up in cubic volumes of comoving side length $L_{\rm box}=150
\,h^{-1}\,\mathrm{Mpc}$ with $256^3$ particles (see
Tab.~\ref{tab:simu_para}). The cosmology chosen
reflects the WMAP5 results \citep{WMAP5} and thus has a lower value
for $\sigma_8$ than the Virgo simulation. It uses the GADGET-2 code to
simulate the evolution of dark matter particles
\citep{2005MNRAS.364.1105S} and has a softening length of $2l_{\rm
  soft} \approx 30 \,h^{-1} \,\mathrm{kpc}$.

\subsection{Ray-tracing}
\label{sec:ray-tracing}

The output of numerical simulations are three-dimensional
distributions of $N_{\rm par}$ particles in cubic boxes over a range
of redshift values. In order to compare the results with the predicted
convergence power spectrum from the halo model, we make use of the
\emph{multiple-lens-plane ray-tracing} algorithm (see e.g.,
\citealt{2000ApJ...530..547J,2008arXiv0809.5035H}) to construct
effective convergence maps. The basic idea is to introduce a series of
lens planes perpendicular to the central line-of-sight of the
observer's backward light cone. In this way, the matter distribution
within the light cone is sliced and can be projected onto the
corresponding lens plane. By computing the deflection of light rays
and its derivatives at each lens plane, one simulates the photon
trajectory from the observer to the source by keeping track of the
distortions of ray bundles. In this way, the continuous deflection of
light rays is approximated by a finite number of deflections at the
lens planes. As a result, one obtains the Jacobian matrix for the lens
mapping from source to observer and can construct convergence maps.

For both simulations, a similar number of around 200 effective
convergence maps were produced, with source galaxies situated at a
single redshift of either $z_{\rm s}=1$ or $z_{\rm s}=2$ (see
Tab.~\ref{tab:simu_para}). The maps produced with the Gems simulation
have an area of 16 $\deg^2$, while the ones from Virgo are much
smaller, with 0.25~$\deg^2$. 

\begin{figure}
    \centering
    \includegraphics[angle=0,width=0.8\textwidth]{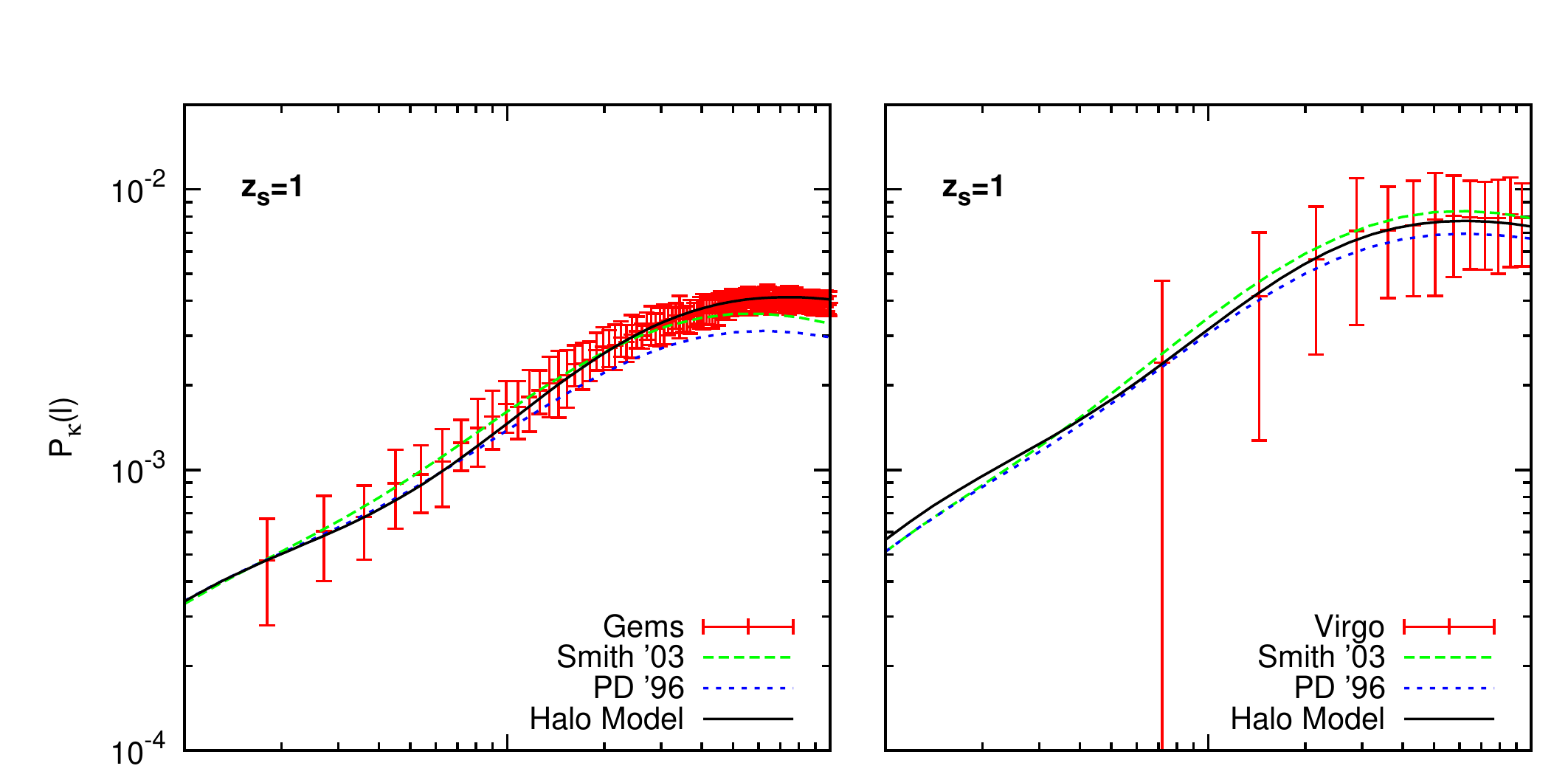}
    \vspace{0.3cm}
    \includegraphics[angle=0,width=0.8\textwidth]{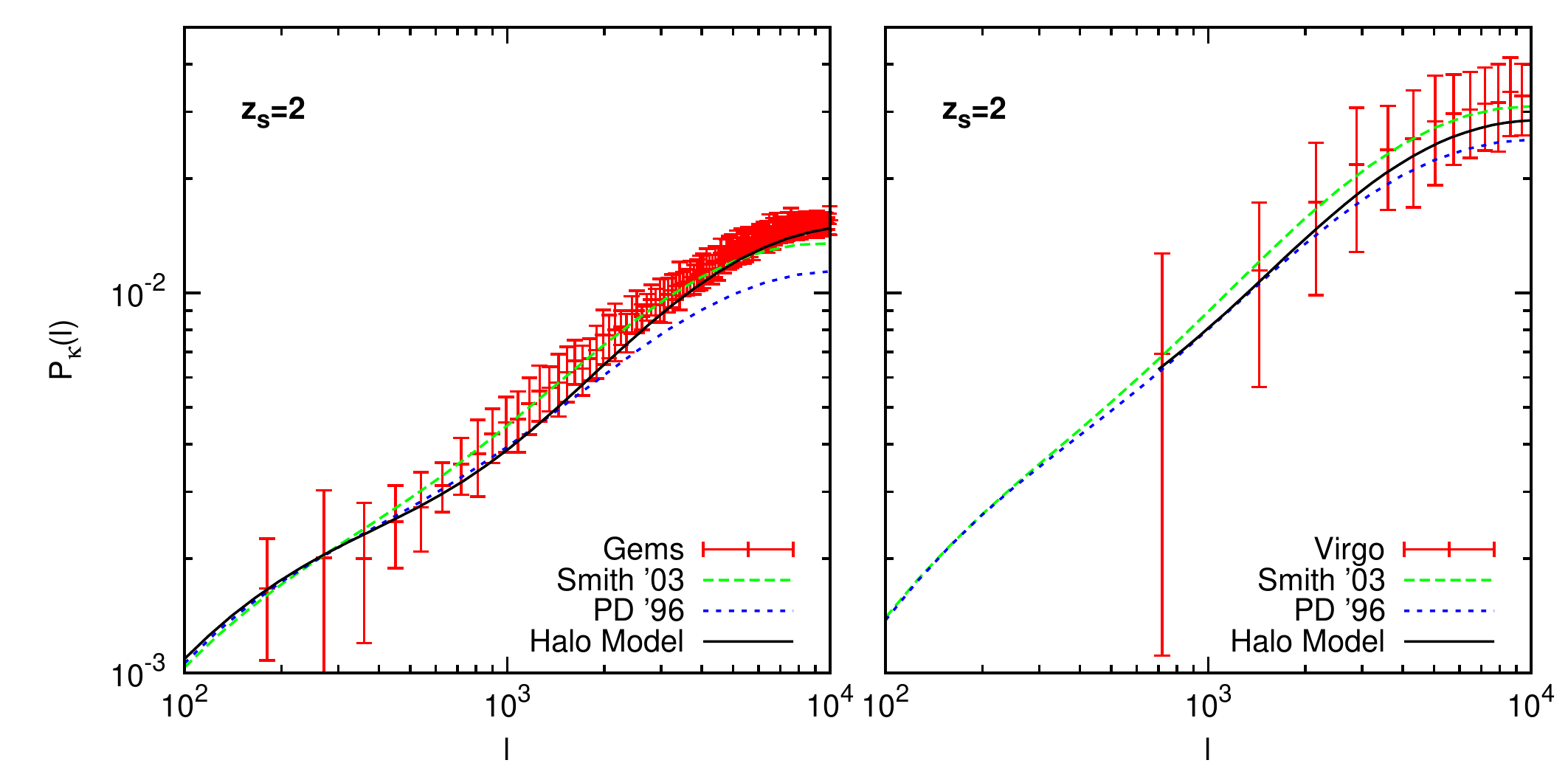}
    \caption{Dimensionless convergence power spectrum ${\cal P}_\kappa(\ell)$
      against wave-number $\ell$ for galaxy sources at redshift $z_{\rm s}=1$
      (upper panels) and $z_{\rm s}=2$ (lower panels). Points with error bars
      show measurements from the two sets of numerical simulations: Gems (left
      plots) and Virgo (right plots). Both simulations use a similar particle
      setup but differ in the area of the $\kappa$-maps (see
      Tabs.~\ref{tab:cosmo_parameters} and \ref{tab:simu_para}). They are
      compared with the corresponding results obtained with fitting formulae
      from Smith et al. (long-dashed line), Peacock-Dodds (short-dashed line),
      and the halo model predictions for a deterministic halo concentration
      (solid line). See text for a discussion.}
    \label{abb:pkappa_virgo}
\end{figure}

\subsection{Convergence power spectrum}
\label{sec:conv-power-spectrum}

To test the accuracy of the halo model approach in describing the
non-linear evolution of dark matter, we compare
the dimensionless projected power spectrum,
computed in the halo model approach, to the ones estimated from the numerical
$N$-body simulations. 

The dimensionless convergence power spectra of the simulations are
measured from the real-space two-dimensional convergence maps of
length $L_{\rm map}$ and grid-size $N_{\rm bin}$. For this, we first
apply a Fast Fourier Transform\footnote{For the FFT we use an
  algorithm from the GNU scientific library (see
  \url{http://www.gnu.org/software/gsl/} for more details).} to obtain
$\tilde{\kappa}(\ell)$ on each grid-point. Then, we estimate the power
spectrum at a wave-number $\ell$ for the $k$-th convergence map by
averaging over all Fourier modes in the band ${\bs \ell}_{\rm b}=\{\bs
\ell \, | \, \ell-\Delta \ell/2 \leq |\bs \ell| \leq \ell+\Delta
\ell/2$\}, i.e.,
\begin{equation}
  \label{eq:estimator_pkappa}
  \hat{\cal P}^{(k)} (\ell)= \rez{N_{\rm p}(\ell)} \sum_{\bs \ell \in
    \bs \ell_{\rm b}}
  \frac{\ell^2}{2\pi} \, |\tilde{\kappa}(\ell)|^2 \,,   
\end{equation}
where $N_{\rm p}(\ell)$ denotes the number of modes in each band
$\ell_{\rm b}$ of width $\Delta \ell$. Finally, the ensemble average
of the dimensionless power spectrum is obtained by averaging over the
results of the various convergence maps, i.e., $\mathcal{P}(\ell) =
\rez{N_{\rm map}} \sum_{k=1}^{N_{\rm map}} \hat{\cal
  P}^{(k)}(\ell_i)$. The error bars for the power spectrum estimate
are computed from the dispersion over the $N_{\rm map}$ convergence
maps used, and are due to sample variance. Since modes with a length
similar to the side length of the convergence map are only poorly
represented, the sampling variance is largest for small $\ell$. The
effect is stronger for the Virgo simulation than for the Gems
simulation, since the Gems convergence maps have a much larger area
$A_{\rm map}$.

Fig.~\ref{abb:pkappa_virgo} shows a good agreement between the halo
model and the $N$-body simulation convergence power spectra. We also
include, in Fig.~\ref{abb:pkappa_virgo}, the convergence power spectra
computed using the fitting formulae of \citet{1996MNRAS.280L..19P} and
\citet{2003MNRAS.341.1311S}. Both, the halo model and the fitting
formulae, show a better agreement with simulations for the lower
source redshift case. On small scales, the halo model and
the Smith et al. fitting formula agree well with the simulations,
whereas the Peacock-Dodds fitting formula has too little power, in
particular in the case of the Gems simulation. On intermediate scales
($\ell \simeq 10^3$), the halo model is less accurate than the Smith
et al. fitting formula, suggesting that the halo model suffers from
the halo exclusion problem on these scales, as described, e.g., in
\citet{2005ApJ...631...41T}. This means that, while in simulations
halos are never separated by distances smaller than the sum of their
virial radii, this is not accounted for in the framework of our halo
model and is probably the cause for the deviation. The good agreement
of the Smith et al. formula is not surprising, since it is based on
simulations of similar resolution than the ones we consider here. In
contrast, a similar comparison using the convergence power spectrum
estimated with the Millennium Run \citep{2008arXiv0809.5035H} clearly
favors the halo model prediction over of the two fitting formulae,
with both fitting formulae strongly underestimating the power on
intermediate and small scales.

The good overall accuracy of the halo model results was expected since its
ingredients, such as the mass function and the halo profile, were obtained
from $N$-body simulations. We note that, in this analysis, we used a
deterministic concentration parameter, since the use of a stochastic one has
only a small effect on the small scales of the convergence power spectrum.

\begin{figure}
   \centering
   \includegraphics[angle=0,width=0.85\textwidth]{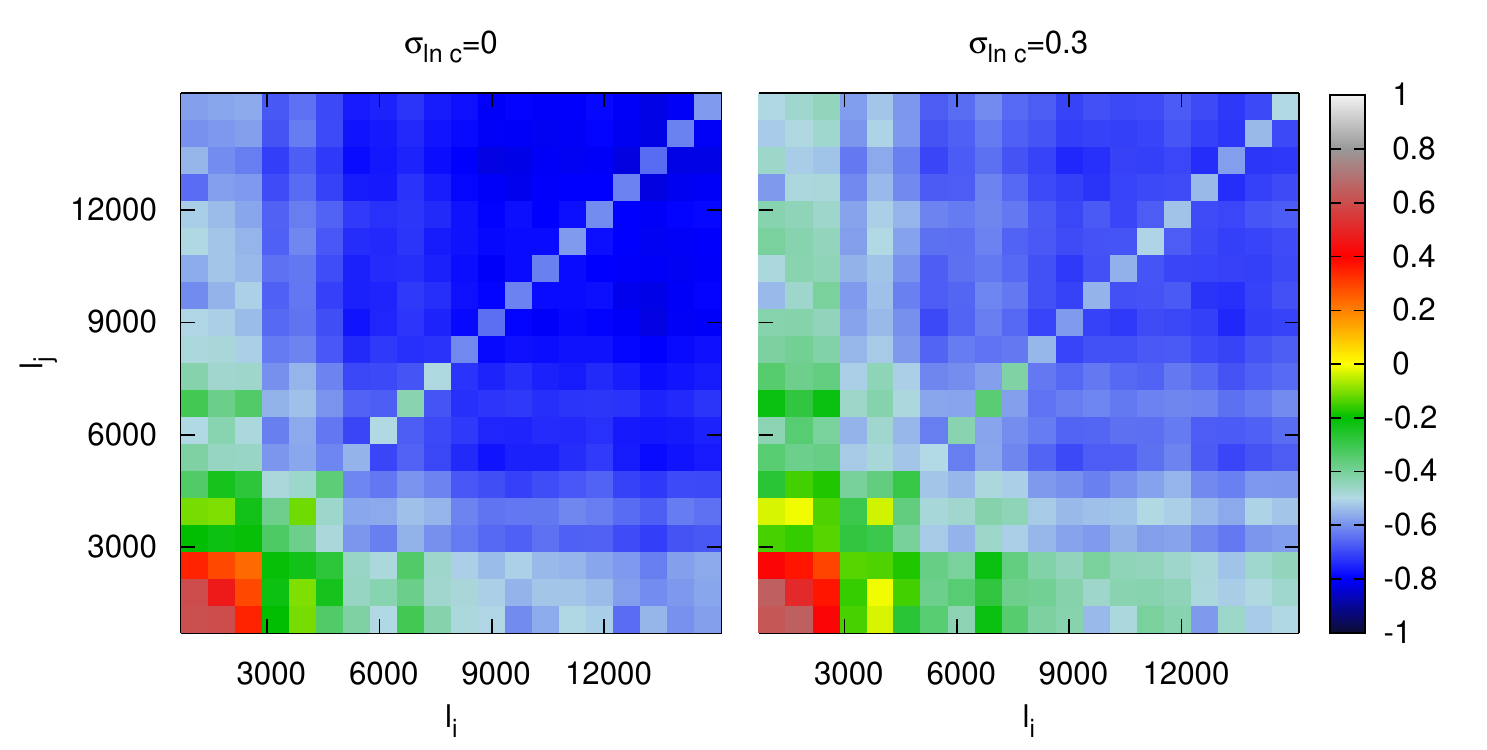}
   \caption{Relative difference $\Delta
     \mathcal{C}_{ij}/\mathcal{C}_{ij}$ (see Eq.~\ref{eq:rel_diff}) between
     the halo model prediction for the covariance of the convergence
     power spectrum and the results from the Virgo
     simulation ($z_{\rm s}=1$) against wave-numbers
     $(\ell_i,\ell_j)$. The binning scheme is linear with a constant
     bin width of $\Delta l=720$ ranging from $\ell_0=720$ to
     $\ell_{20}=15120$. The left panel displays the full covariance,
     including all four halo terms for the trispectrum, and uses a
     deterministic concentration-mass relation denoted by $\sigma_{\ln
       c}=0$. The right panel illustrates the same covariance but with
     a stochastic concentration distribution of width $\sigma_{\ln
       c}=0.3$ for the 1-halo term of the trispectrum.}
   \label{fig:cov_virgo_zs1}
\end{figure}

\begin{figure}
   \centering
   \includegraphics[angle=0,width=0.9\textwidth]{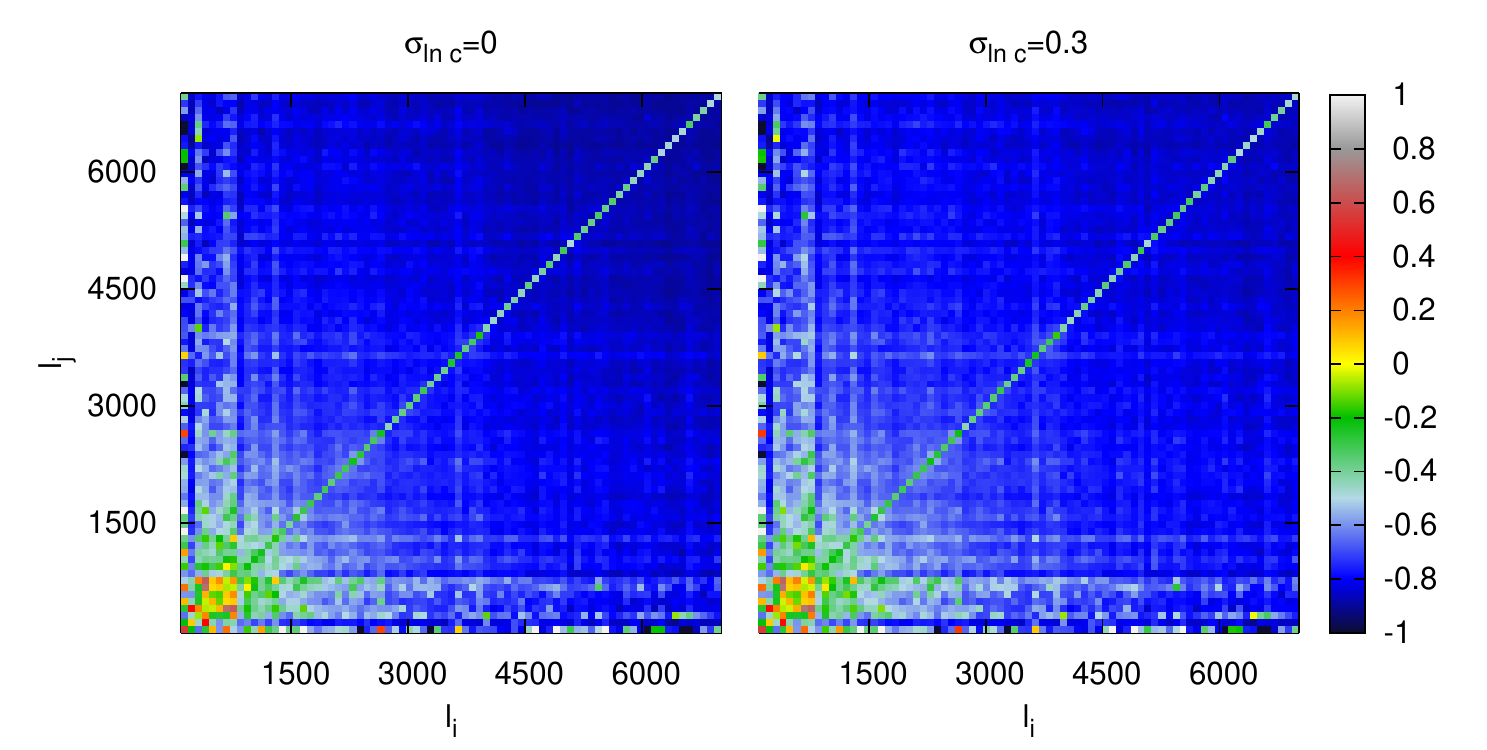}
   \caption{Relative difference $\Delta
     \mathcal{C}_{ij}/\mathcal{C}_{ij}$ (see Eq.~\ref{eq:rel_diff}) between
     the halo model prediction for the covariance of the convergence
     power spectrum and the results from the Gems
     simulation ($z_{\rm s}=1$) against wave-numbers
     $(\ell_i,\ell_j)$. The binning scheme is linear with a constant
     bin width of $\Delta l=90$ ranging from $\ell_0=90$ to
     $\ell_{20}=7020$. The left panel displays the full covariance,
     including all four halo terms for the trispectrum, and uses a
     deterministic concentration-mass relation denoted by $\sigma_{\ln
       c}=0$. The right panel illustrates the same covariance but with
     a stochastic concentration distribution of width $\sigma_{\ln
       c}=0.3$ for the 1-halo term of the trispectrum.}
   \label{fig:cov_gems_zs1}
\end{figure}

\subsection{Covariance of the convergence power spectrum }
\label{sec:proj-power-spec_cov}

The similarity between simulation and halo model power spectra
was to some extent expected. The ability to make an accurate description
 of higher-order
correlations provides a stronger test of the halo model.
Due to its
important role in calculating the error of the power spectrum and for
parameter estimates, we focus in this section
on the accuracy of the covariance of the
dimensionless convergence power spectrum.

We need an appropriate estimator for the power spectrum covariance of the
simulations. 
As each simulation provides $N_{\rm map}$ different
$\kappa$-maps, we have $N_{\rm map}$ realizations of the power
spectrum. From these we estimate the covariance by applying the
unbiased sample covariance estimator which has for our purpose the
form:
\begin{equation}
\label{eq:cov_est}
C_{\rm sim}(\ell_i,\ell_j) = \rez{N_{\rm map}-1} \left[ \sum_{k=1}^{N_{\rm map}} \hat{\cal
    P}_\kappa^{(k)}(\ell_i) \hat{\cal P}_\kappa^{(k)}(\ell_j) -
  \rez{N_{\rm map}} \left(\sum_{k=1}^{N_{\rm map}} \hat{\cal P}_\kappa^{(k)}(\ell_i)\right) \left(\sum_{k=1}^{N_{\rm map}} \hat{\cal P}_\kappa^{(k)}(\ell_j) \right)\right] \,,
\end{equation}
where $\hat{\cal P}_\kappa^{(k)}(\ell_i)$ is the projected power
spectrum estimate of the $k$-th effective convergence map at a
wave-number $\ell_i$ (see Eq.~\ref{eq:estimator_pkappa}). We evaluate
$C_{\rm sim}$ for both, Virgo and Gems simulations, for the case
$z_{\rm s}=1$.

The halo model results, $\mathcal{C}_{\rm halo}$, are calculated as
described in Sect.~\ref{sec:halo-model} and include all terms of the
non-Gaussian covariance. In their computation, we use our fiducial
halo model with the ingredients summarized in
Sect.~\ref{sec:ingredients}, and use the cosmological parameters
values corresponding to each simulation, given in
Tab.~\ref{tab:cosmo_parameters}, for the case $z_{\rm s}=1$. We
consider both, a deterministic concentration-mass relation and a
stochastic one, with dispersion $\sigma_{\ln c}=0.3$ for the 1-halo
term of the trispectrum.

We compare the halo model with the simulations covariance matrices 
considering their relative deviation
\begin{equation}
\label{eq:rel_diff}
\left(\frac{\Delta \mathcal{C}_{ij}}{\mathcal{C}_{ij}}\right) =
\frac{\mathcal{C}_{\rm halo}(\ell_i,\ell_j) -\mathcal{C}_{\rm sim}(\ell_i,\ell_j)}{\mathcal{C}_{\rm sim}(\ell_i,\ell_j)} \,.
\end{equation}
We note that although the number of available convergence maps for our
simulations ($\sim 200$) is too small to obtain a percentage level
accuracy for the covariance estimate \citep{2009arXiv0902.0371T}, we assume
that the covariance from the simulations is the reference one, and put
it in the denominator of Eq.~(\ref{eq:rel_diff}). The comparisons are
displayed in Figs.~\ref{fig:cov_virgo_zs1} and \ref{fig:cov_gems_zs1}
for the Virgo and Gems simulation, respectively. In the case of the
Virgo simulation with $\sigma_{\ln c}=0$, the best agreement is a
relative deviation of $20\%-40\%$ found on intermediate scales between
$\ell \simeq 1500$ and $\ell \simeq 5000$. For small scales $(\ell
\gtrsim 6000)$ the halo model underestimates the simulation by $60\%$
or more. On large scales, i.e., for wave-numbers $\ell \lesssim 1500$,
the halo model overestimates the simulation by around $60\%$. However,
since the simulation covariances suffer from a large sampling variance
due to the small size of the convergence maps, the comparison is not
meaningful on these scales. The agreement improves for $\sigma_{\ln
  c}=0.3$, in this case there is a larger range of scales where the
deviation is small. For the Gems simulation the agreement is much
better. There is now an interval of scales ($\ell\lesssim 1500$) where
the deviation is in the range $0\%-40\%$. Throughout all the scales
probed, the deviation on the diagonal of the covariances is around
$20\%-40\%$. However, for off-diagonal components, at small scales,
the agreement is still poor. Finally, the improvement of considering
$\sigma_{\ln c}=0.3$ is less significant than in the Virgo case.

Although the halo model predictions strongly deviate from the
simulation estimates of the covariance on small scales, this analysis
does not necessarily imply a poor accuracy of those predictions.
Indeed, the simulation covariances estimated in this analysis have a
strong scatter. In particular, in the case of the Gems simulation, we
found from bootstrap subsamples of 50 convergence maps that the
resulting covariances can deviate up to $20\%$ from the average
covariance of the complete sample \citep{pielorz:2008}. This is in
agreement with the results by \citet{2009arXiv0902.0371T} who need to use
5000 simulations to obtain an estimate of the matter power spectrum
covariance at a sub-percent level accuracy.

There are however very recent indications that Eq.~(\ref{eq:wlcov}) indeed
underestimates the covariance of the convergence power spectrum on small
scales \citep{2009arXiv0906.2237S}. In particular, sample variance in the
number of halos in a finite field is not accounted for by
Eq.~(\ref{eq:wlcov}). Indeed, the mass function yields a mean number density,
but there are fluctuations on the number of halos due to the large-scale mass
fluctuations. Sample variance in the number of clusters in a volume-limited
survey \citep{2003ApJ...584..702H} has been considered in cluster abundance
studies \citep[e.g.,][]{2009ApJ...692.1033V}. In the halo model framework, the
sample variance in the number of halos was derived in
\cite{2007NJPh....9..446T} and its contribution to the covariance of the
convergence power spectrum was found, in \cite{2009arXiv0906.2237S}, to boost
the non-Gaussian errors of a 25 $\deg^2$ survey by one order of magnitude on
scales $\ell \approx 10^4$. The increase is reduced for larger survey areas.

\section{Fitting formula for the covariance of the convergence power spectrum}
\label{sec:fitting-formula}

\begin{figure}
   \centering
   \includegraphics[angle=0,width=0.55\textwidth]{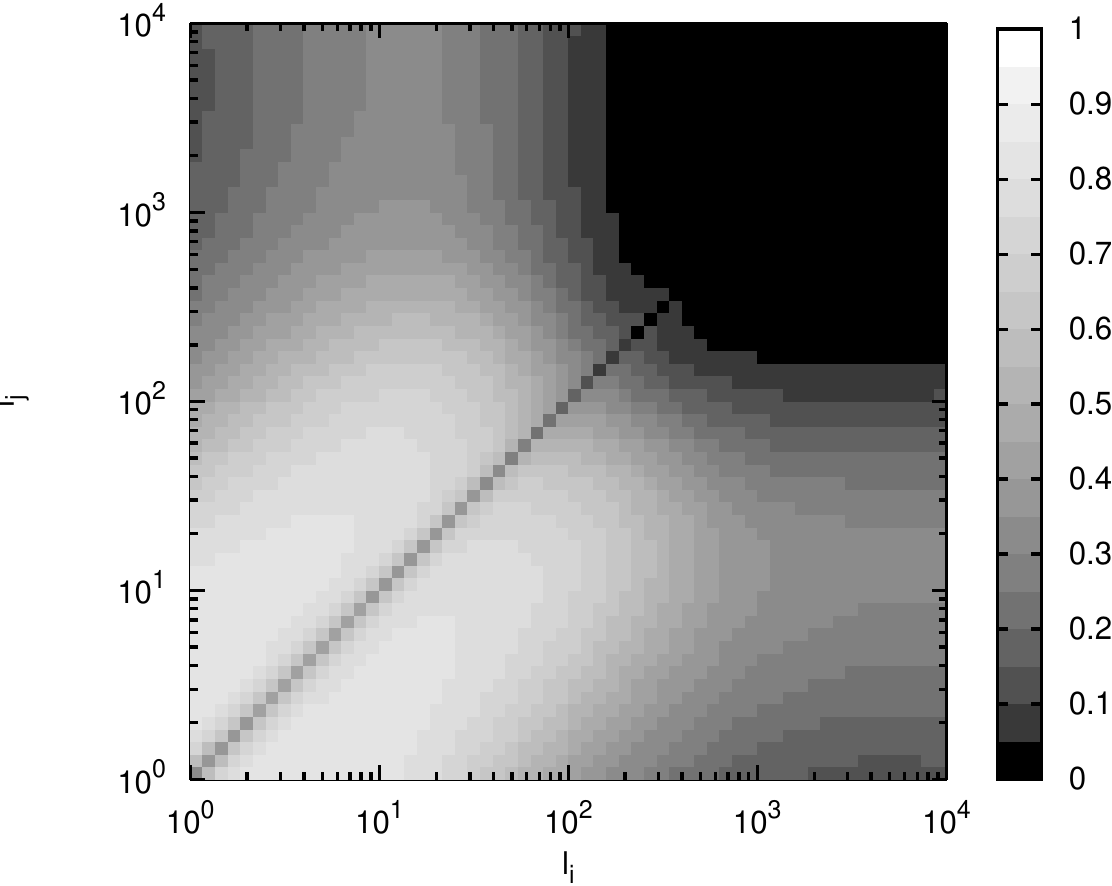}
   \caption{Ratio ${\cal C}_{\rm pt}/{\cal C}^{\rm NG}$ against
     wave-numbers $(\ell_i,\ell_j)$. The covariance in tree-level
     perturbation theory resembles the non-Gaussian covariance on very
     large scales (i.e., $\ell_i=\ell_j\leq 20$) to approximately
     $80\%-90\%$ except along the diagonal and the closest
     off-diagonals. Here a good approximation of the non-Gaussian
     covariance requires at least one additional halo term. For
     wave-numbers $\ell_i=\ell_j\geq 100$, an accurate description of
     the non-Gaussian covariance requires at least the 1-halo term.}
   \label{fig:cov_pt}
\end{figure}

\begin{table}
  \renewcommand{\arraystretch}{1.2}
  \caption{ The fiducial model used for the fitting procedure. See text for details.}
\label{tab:fiducial_model}
\centering
\begin{tabular}[h]{c c c c c c c c }
  \hline\hline
  $\Omega_{\rm m}$ & $\Omega_\Lambda$ & $h$ &
  $\Omega_{\rm b}$ & $\sigma_8$ &
  $n_{\rm s}$ & $\sigma_{\ln c}$ & $T(k)$ \\ \hline 
  0.28 & 0.72 & 0.73 & 0.045 & 0.82 & 1.0 & 0.3 
  & EH \\ \hline
\end{tabular}
\end{table}

Future weak lensing surveys will provide much more precise measurements of the
convergence power spectrum. In order to obtain robust constraints on
cosmological parameters, accurate estimates of both the power spectrum and its
covariance are needed.

\subsection{Methodology}

The number of measured power spectra is, in general, insufficient to infer
the complete covariance directly from observations. One has thus to derive
it either from ray-tracing maps of numerical $N$-body simulations or
with an analytic approach. A drawback of the first method is that it
requires a large number of realizations and, in addition, is very
time-consuming if an exploration of the covariance in the parameter
space is needed.

In the previous sections, we derived the covariance, and in particular
its non-Gaussian part, with an analytic approach. This computation is,
however, time-consuming and it might be useful to obtain an accurate
covariance in a faster way. A first approach would be to rely on
stronger approximations. For example, a commonly used approximation
consists on evaluating the non-Gaussian covariance from $T_\kappa
\approx T_{\rm pt} + T_{\rm 1h}$, instead of using the full
trispectrum. We saw in Sect.~\ref{sec:halo-model} that this
approximation recovers the full trispectrum for scales $\ell \gtrsim
10^3$ but deviates by $\sim 40\%$ on scales $\ell \lesssim 10^2$ for
square configurations (compare also with Fig.~\ref{fig:cov_pt}).

An alternative approach that we consider in the following, is to find
a fitting formula for the halo model covariance that can be
subsequently used without the need for implementing the halo model. We
will provide a fit only for the non-Gaussian part of the halo model
covariance, since the Gaussian part only depends on the non-linear
convergence power spectrum and can thus be accurately computed without
relying on the halo model. The inclusion of non-Gaussian errors
increases the total covariance and one might think of fitting the
ratio between the non-Gaussian and Gaussian terms. However, this is
not a good quantity to fit, since the Gaussian contribution is
diagonal and binning-dependent. In contrast, a similar ratio was
fitted in the real space, where the Gaussian term is non-diagonal and
binning-independent, using a non-Gaussian contribution measured from
$N$-body simulations \citep{2007MNRAS.375L...6S}. In Fourier space,
there is some relevant analogous information contained in the
tree-level perturbation theory trispectrum. Indeed, pursuing the
analogy with the real-space fit, $T_{\rm pt}$ is a non-diagonal and
binning-independent quantity, with a lower amplitude than the full
trispectrum, approaching it at large scales. We thus compute ${\cal
  C}_{\rm pt}$, the covariance predicted in tree-level perturbation
theory on large scales (see Eq.~\ref{eq:cov_perturbation} and
Appendix~\ref{sec:PT} for a detailed derivation) and compare it with
our covariance computed with the full halo model trispectrum, ${\cal
  C}^{\rm NG}$, defined in Eq.~(\ref{eq:wlcov}). The ratio between the
two covariances is shown in Fig.~\ref{fig:cov_pt}. In agreement with
Fig.~\ref{fig:trikappa}, the covariance predicted by tree-level
perturbation theory contributes only $\sim 50\%$ to the complete
non-Gaussian covariance along the diagonal. On very large scales
$(\ell<100)$, the ratio ${\cal C}^{\rm NG}/{\cal C}_{\rm pt}$ lies
between 1.1 and 2. On smaller scales, ${\cal C}_{\rm pt}$ decreases
fast and is no longer useful for fitting purposes.

This discussion motivates us to use two different fitting formulae to
model the non-Gaussian covariance over the whole range of scales. On
large scales ($1<\ell<200$), we model the ratio ${\cal C}^{\rm
  NG}/{\cal C}_{\rm pt}$ as a polynomial in the wave-numbers,
$\ell_i\,,\ell_j$, and the dimensionless non-linear convergence power
spectrum $\mathcal{P}_\kappa(\ell)$. More precisely, we assume
\begin{equation}
  \label{eq:cov_large}
  \mathcal{C}_< (\ell_i,\ell_j) = \mathcal{C}_{\rm pt}(\ell_i,\ell_j) \left[
    1.1 + a_0 \ell_{\rm min} + a_1 \ell_{\rm max} + a_2
    {\mathcal{P}_\kappa(\ell_{\rm min})} +
    a_3 {\mathcal{P}_\kappa(\ell_{\rm max})} \right] \,,
\end{equation}
where $\ell_{\rm min}= \min(\ell_i, \ell_j)$ and $\ell_{\rm
 max}=\max(\ell_i, \ell_j)$ which ensures the symmetry property
$\mathcal{C}_{<}(\ell_i,\ell_j)=\mathcal{C}_{<}(\ell_j,\ell_i)$. In
this way, $\mathcal{C}_{<} \rightarrow 1.1 \, \mathcal{C}_{\rm pt}$
for large scales as predicted by the halo model (see
Fig.~\ref{fig:cov_pt}), whereas the non-linear clustering on smaller
scales is encoded in the first-order polynomial in wave-numbers and
power spectra. On small scales, i.e. for $200<\ell<8000$, we model
directly the non-Gaussian covariance amplitude using a second-order
polynomial in the power spectrum, such that
\begin{equation}
  \label{eq:cov_small}  
  \mathcal{C}_{>}(\ell_i,\ell_j) = a_4{\mathcal{P}_\kappa(\ell_{\rm min})} +
  a_5{\mathcal{P}_\kappa(\ell_{\rm max})} + a_6
  {\mathcal{P}_\kappa^2(\ell_{\rm min})} +
  a_7{\mathcal{P}_\kappa^2(\ell_{\rm max})} + a_8 {\mathcal{P}_\kappa(\ell_{\rm min})}{\mathcal{P}(\ell_{\rm max})} \,.
\end{equation}
To ensure a smooth transition from small to large wave-numbers, we
consider a linear combination of the two matrices defined in
Eqs.~\eqref{eq:cov_large} and \eqref{eq:cov_small} with weightings of
third-order in the wave-numbers, such that the full non-Gaussian
covariance becomes
\begin{equation}
  \label{eq:fitting_formula}
  \mathcal{C}^{\rm NG}_{\rm fit}(\ell_i,\ell_j) =
  \frac{\ell_0^3}{\ell_0^3+(\ell_i + \ell_j)^3} \;
  \mathcal{C}_{<}(\ell_i,\ell_j) + \frac{(\ell_i +
    \ell_j)^3}{\ell_0^3+(\ell_i + \ell_j)^3} \;
  \mathcal{C}_{>}(\ell_i,\ell_j)  \,, 
\end{equation}
with the transition scale fixed at $\ell_0=600$. We use thus a model
with nine free parameters, which is a compromise between expressive
power and simplicity, and perform the fit using the range $1 \leq \ell
\leq 8000$. More precisely, we find the coefficients for the final fit
formula by performing two least square fits to
Eq.~(\ref{eq:cov_large}) and Eq.~(\ref{eq:cov_small}), respectively.
To obtain the coefficients for small (large) scales we computed the
corresponding covariance as described above within the halo model
approach with 61 bins in the range $10^2 \leq \ell \leq 10^4$ ($1 \leq
\ell \leq 10^3$) and employed the condition $\ell_i +\ell_j \geq 200$
($\ell_i +\ell_j < 200$) for the fit.

The covariances used in the fitting procedure, both tree-level
perturbation theory and halo model covariance, depend on perturbation
theory polyspectra. These were evaluated from the expressions derived
in Appendix~\ref{sec:PT} using a linear matter power spectrum computed
with the Eisenstein-Hu transfer function \citep{1998ApJ...496..605E}
and assuming the WMAP5-like fiducial model shown in
Tab.~\ref{tab:fiducial_model}. The non-linear convergence power
spectrum, used in the polynomial expressions, was evaluated from the
same linear power spectrum. In addition, the halo model non-Gaussian
covariance was evaluated using the input parameters as described in
Sect.~\ref{sec:ingredients}, including a stochastic concentration-mass
relation with $\sigma_{\ln c}=0.3$ for the 1-halo term of the
trispectrum.

\begin{table}
  \renewcommand{\arraystretch}{1.2}
  \caption{Best-fit values for the parameters $(p_0,p_1,p_2)$ of the
    redshift-fit for the set of coefficients $(a_0,\ldots,a_8)$. The 3-parameter
    fit is defined in Eq.~(\ref{eq:fitaz}) and is valid in the range $z_{\rm s} \in[0.5,2]$.}

\label{tab:fitaz}
\centering
\begin{tabular}[b]{c c c c}
  \hline\hline
  & $p_0$ & $p_1$ & $p_2$ \\
  \hline 
  $a_0$       & $4.974\times 10^{-2}$ & $0.8385$ & $-7.877\times 10^{-3}$  \\
  $a_1$     & $1.854\times 10^{-3}$ & $2.614$  &  $5.926\times 10^{-3}$   \\
  $a_2$ & $-2.037\times 10^{5}$ & $2.350$  & $7.962\times 10^{3}$ \\
  $a_3$ & $-5.564\times 10^{4}$ & $3.152$  &  $-1.625\times 10^4$   \\
  $a_4$     & $3.149\times 10^{-6}$ & $0.1108$ & $-2.172\times 10^{-6}$   \\
  $a_5$     & $5.349\times 10^{-8}$ & $1.332$  & $-9.222\times 10^{-8}$   \\
  $a_6$    & $-2.406\times 10^{-3}$ & $3.116$  &  $-6.171\times 10^{-4}$  \\
  $a_7$    & $-1.444\times 10^{-4}$ & $5.068$  & $9.201\times 10^{-5}$  \\
  $a_8$    & $-4.285\times 10^{-3}$ & $2.529$  & $2.150\times 10^{-4}$  \\
 \hline
\end{tabular}
\end{table}

\begin{figure}
  \centering
  \includegraphics[angle=0,width=0.49\textwidth]{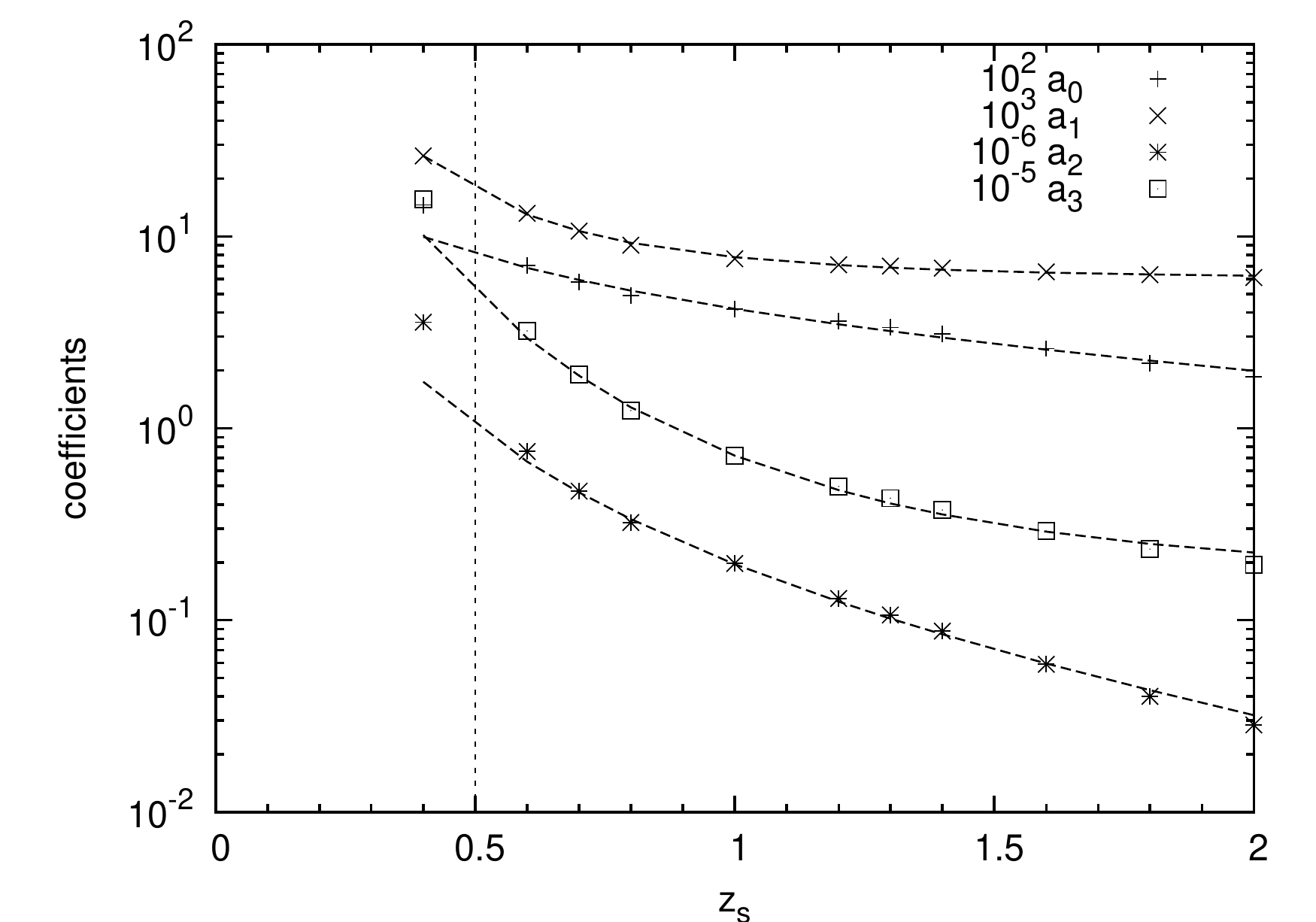}
  \includegraphics[angle=0,width=0.49\textwidth]{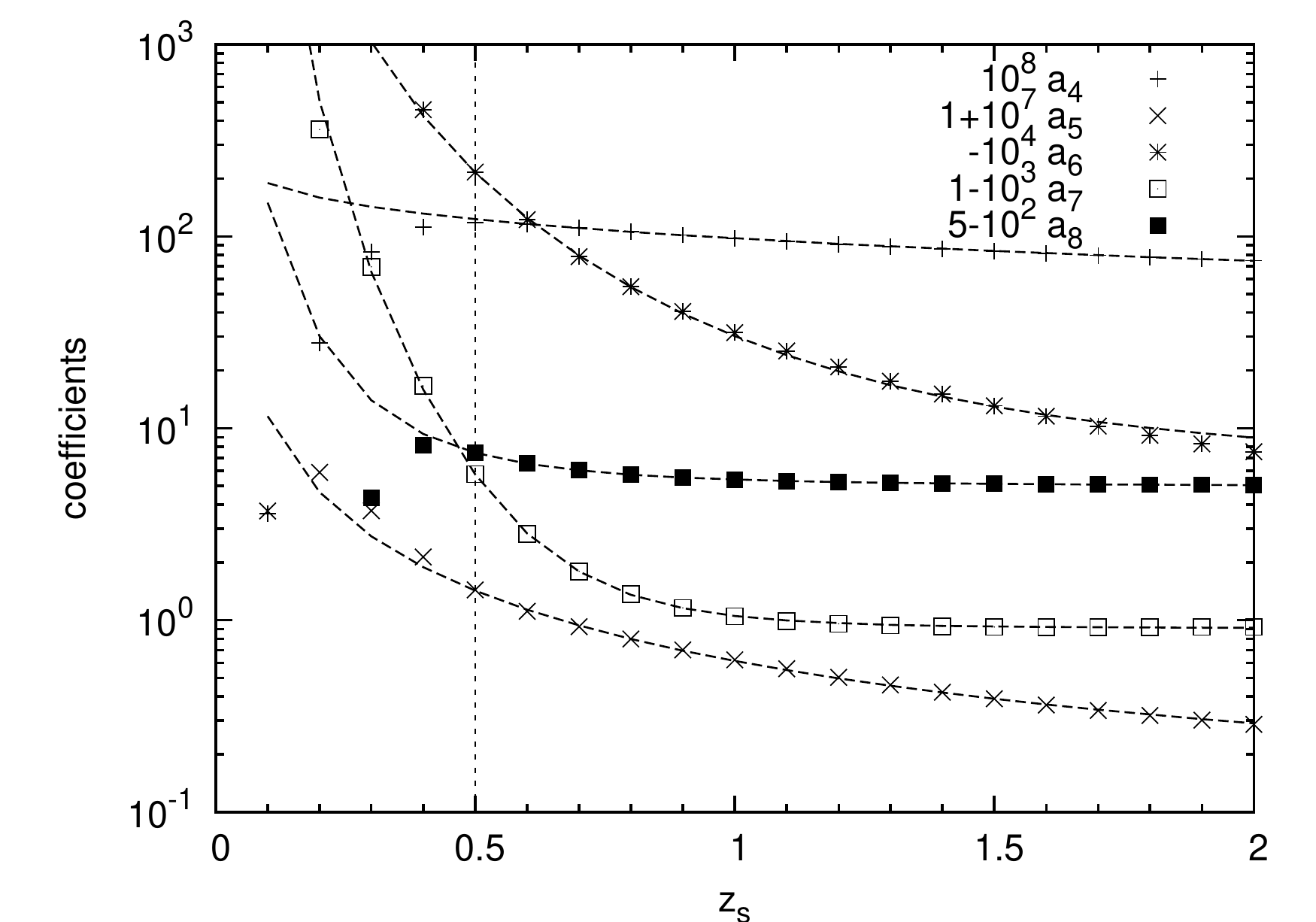}
  \caption{Coefficients $a_i$ of the fitting formula obtained for
    various source redshifts. Left (right) panel shows the
    coefficients of the large (small) scales fitting formula. For
    convenience, scaled versions of the coefficients are shown, as
    indicated in the key. Note for example that $a_5$, $a_7$ and $a_8$
    change sign. The lines show the redshift fit,
    Eq.~(\ref{eq:fitaz}), with the parameters given in
    Tab.~\ref{tab:fitaz}. They provide a good fit for $z_{\rm s}>0.5$
    (indicated by the vertical line).}
  \label{fig:fitaz}
\end{figure}

\subsection{Redshift dependence}

All quantities, $\mathcal{C}^{\rm NG}$, $\mathcal{C}_{\rm pt}$, and
$\mathcal{P}_{\kappa}$, were evaluated for several values of source redshifts
(assuming a single source redshift plane), in the range $0.1 \le z_{\rm s} \le
2.0$. For each redshift, we perform the fit and find a set of best-fit values
for the coefficients $(a_0,\ldots,a_8)$. Next, we fit the best-fit values of
each coefficient as a function of redshift. Some of the best-fit values are
increasing functions of the source redshift, while others decrease with
redshift, and others still are non-monotonic. They all are, however, monotonic
in the redshift range of interest for current and future weak lensing surveys,
$0.5 \leq z_{\rm s} \leq 2.0$. In this range, the best-fit values $a_i(z_{\rm
  s})$ are accurately fitted with
\begin{equation}
  \label{eq:fitaz}
  a(z_{\rm{s}})=\frac{p_0}{(z_{\rm{s}})^{p_1}}+p_2\,.
\end{equation}

Tab.~\ref{tab:fitaz} shows the resulting values of the 27 parameters, which
define our fitting formula for the halo model covariance of the convergence
power spectrum. The behavior of the coefficients with redshift is shown in
Fig.~\ref{fig:fitaz}. We note that a few of them change sign, with most
remaining always positive. In addition, their amplitudes may be quite
distinct, since they are applied to quantities of quite different amplitudes,
such as wave-number, power spectrum or power spectrum squared. The
coefficients of the fitting formula for large scales,
Eq.~(\ref{eq:cov_large}), all decrease with redshift in absolute values. This
compensates the increase of the power spectrum with redshift, allowing for a
decrease of the ratio $\mathcal{C}^{\rm NG}/\mathcal{C}_{\rm pt}$ with
redshift, as expected.

\subsection{Accuracy of the fitting formula}

\begin{figure}
  \centering
  \includegraphics[angle=0,width=0.49\textwidth]{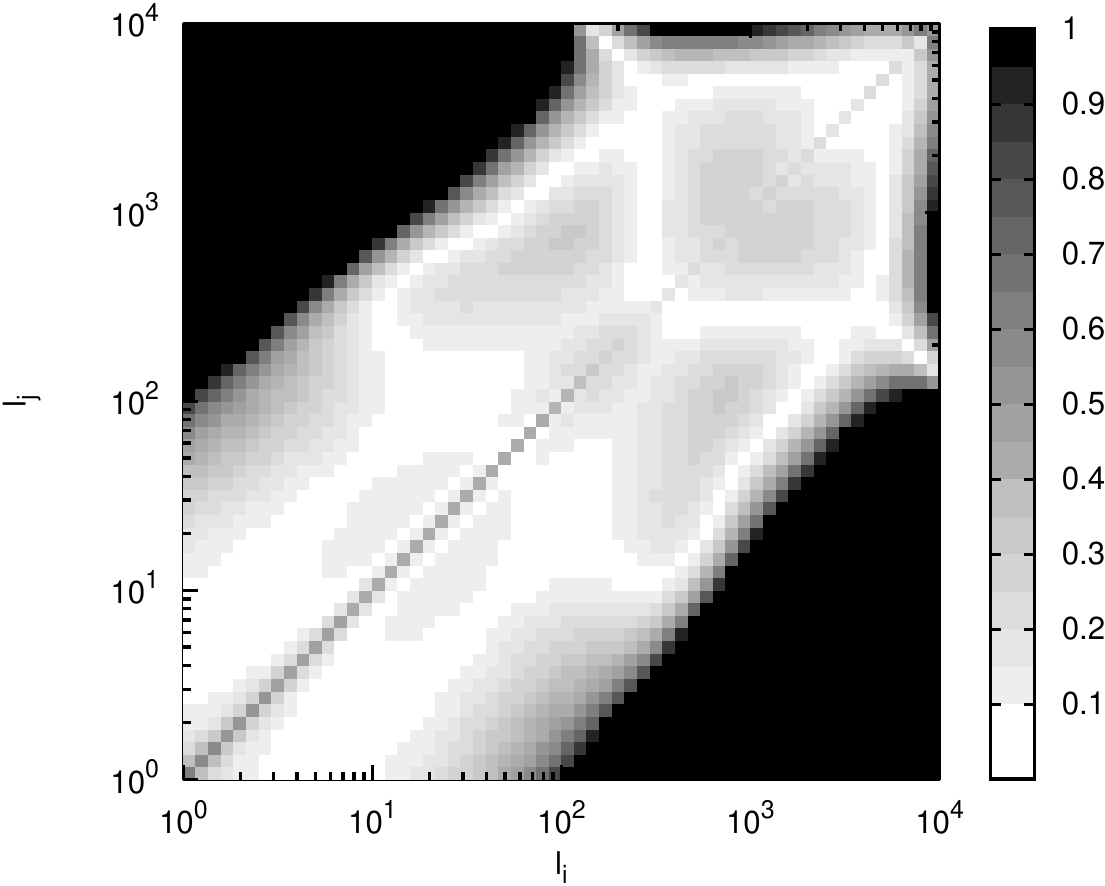}
  \includegraphics[angle=0,width=0.49\textwidth]{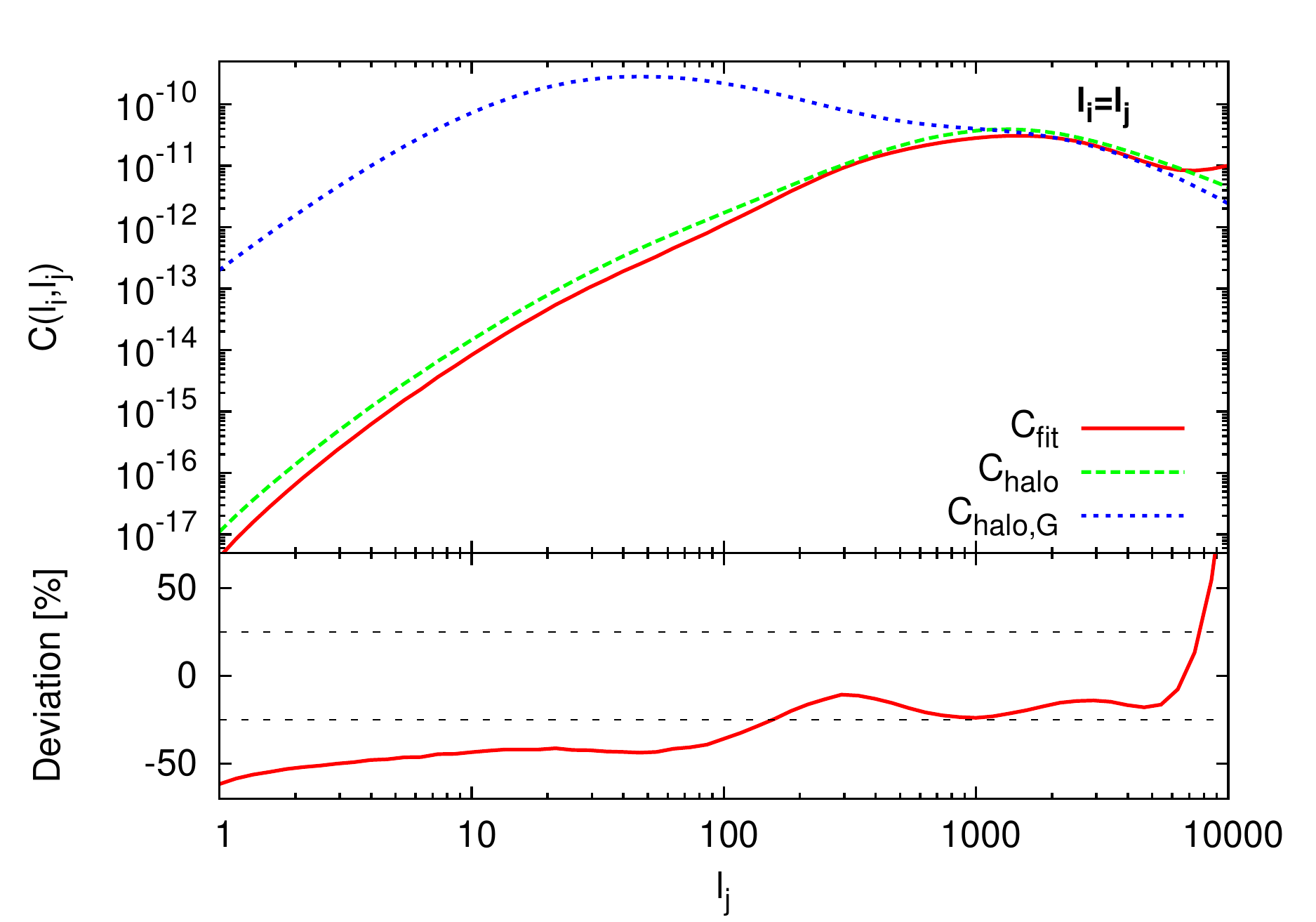}
  \includegraphics[angle=0,width=0.49\textwidth]{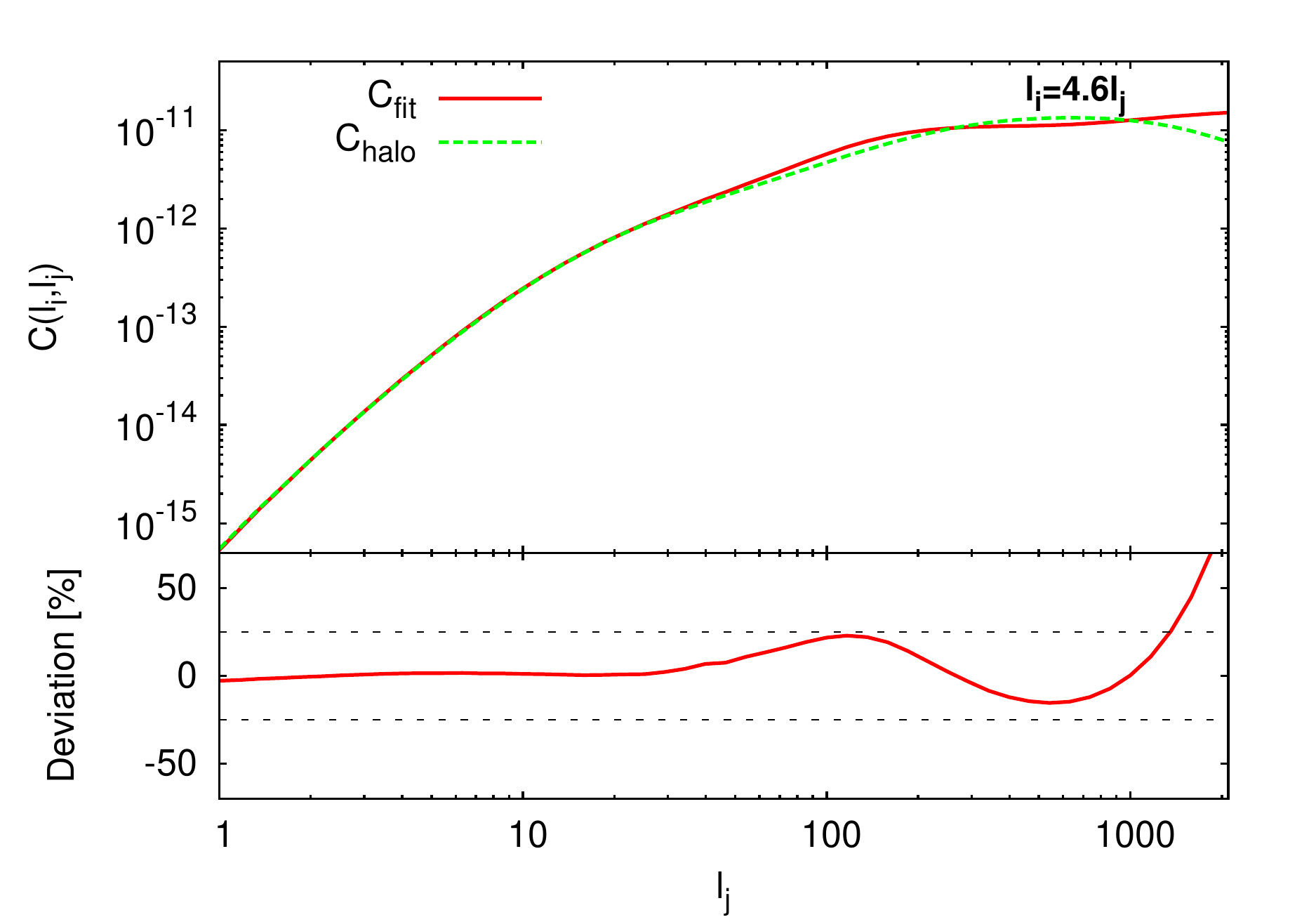}
  \includegraphics[angle=0,width=0.49\textwidth]{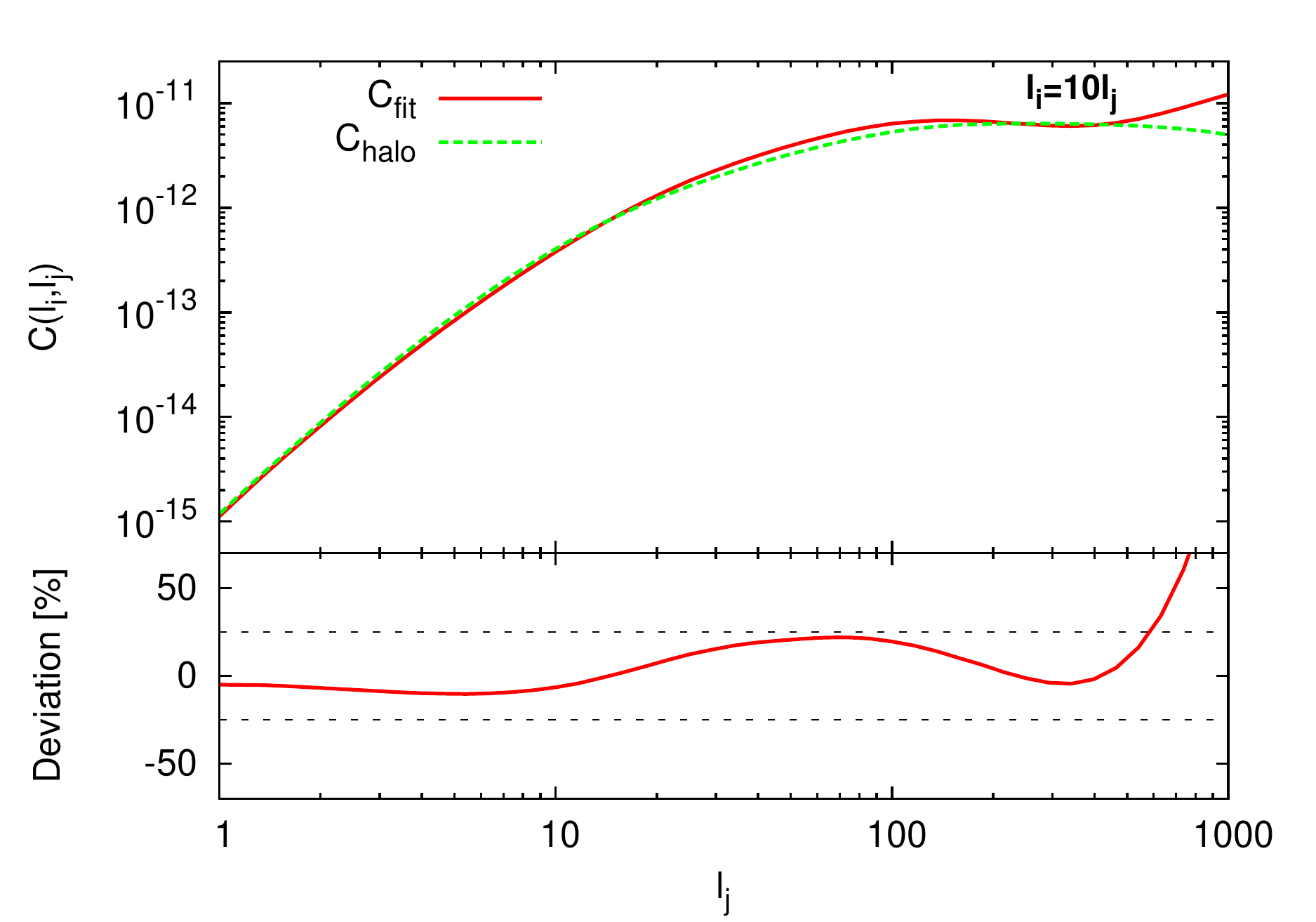}
  \caption{Accuracy of the fitting formula for $z_{\rm s}=1$. Upper
    left panel: relative deviation $\Delta
    \mathcal{C}_{ij}/\mathcal{C}_{ij}$, defined in
    Eq.~(\ref{eq:fitting_error}). Other panels: diagonal cuts through
    the fitted non-Gaussian covariance, showing the covariance on the
    diagonal $\ell_{i}=\ell_{j}$ (upper-right panel), for $\ell_i =
    4.6 \ell_j$ (lower-left) and for $\ell_i = 10 \ell_j$
    (lower-right). All three panels show the fitted and the halo model
    non-Gaussian covariances (as well as the Gaussian for the
    upper-right panel) as function of the lowest scale $\ell_j$ in the
    upper part, and the deviation in percent in the lower part, where
    the dashed lines mark the $25\%$ level. For the Gaussian
    covariance we employed a logarithmic binning with 61 bins in a
    range from $\ell_{\rm low}=1$ to $\ell_{\rm up}=10^5$.}
  \label{fig:ratio_fid}
\end{figure}

 To test the performance of the fitting, we calculate
the relative deviation between the non-Gaussian covariance computed with the
fitting formula and the halo model one. The deviation,
\begin{equation}
  \label{eq:fitting_error}
  \left(\frac{\Delta \mathcal{C}_{ij}}{\mathcal{C}_{ij}}\right) \equiv \frac{\mathcal{C}^{\rm NG}_{\rm
      fit}(\ell_i,\ell_j) - \mathcal{C}^{\rm NG}_{\rm halo}(\ell_i,\ell_j)}{\mathcal{C}^{\rm NG}_{\rm halo}(\ell_i,\ell_j)}\,,
\end{equation}
is computed for every wave-number pair $(\ell_i,\ell_j)$ and shown in 
Fig.~\ref{fig:ratio_fid}, for the case of $z_{\rm s}=1$. 
The upper left panel of Fig.~\ref{fig:ratio_fid} shows the absolute value of
the deviation for each element of the
covariance matrix, while the other panels show the deviation along cuts
through diagonals of the covariance.

The fit works quite well on the off-diagonal elements that are close to the
diagonal, showing an average overestimation of $10\%$. When moving along any
off-diagonal, from larger to smaller scales, 
(lower panels of Fig.~\ref{fig:ratio_fid}), we move from the
first fit, Eq.~(\ref{eq:cov_large}), where the deviation is mostly positive,
to the second one, Eq.~(\ref{eq:cov_small}), where the deviation is mostly
negative. The transition occurs at the local maximum, which indicates that
none of the fitting formulae 
should be extrapolated to the other region. The fits break down on the
smallest scales, with the deviation 
increasing very rapidly when the largest scale reaches $\ell_i \approx 5000$. 

As we move away from the diagonal the fit gets increasingly worse, in
particular the deviations are larger than $100\%$ in the region shown in black
in the upper left panel of Fig.~\ref{fig:ratio_fid}, which correlates very
small with very large scales. The reason for this is that this region was
effectively not fitted, since it is not contained in any of the two blocks
fitted by Eq.~(\ref{eq:fitting_formula}). Restricting to the range where both
scales are between $50 < \ell < 5000$, roughly $90\%$ of the elements show
deviations between $-25\%$ and $+25\%$, with the average of the absolute
deviations being $10\%$. This range contains also $9\%$ of outliers where the
deviations are larger than $\pm 25\%$. The outliers
 occur in the low-amplitude correlations between the largest $(50
< \ell < 200)$ and smallest $(3000 < \ell < 5000)$ scales.

The fit is worse on the diagonal than on the first off-diagonals (where
$\ell_{i}/\ell_{j} < 50$). On the diagonal, the fit always underestimates the covariance. Scales in the
range $50 < \ell < 5000$ show a deviation between $-40\%$ and $-10\%$, with an
average of $-20\%$. The accuracy degrades at larger scales, which is not a
problem since for $\ell < 50$ the non-Gaussian contribution to the total
covariance is negligible, as seen in Fig.~\ref{fig:ratio_fid} (upper right
panel). Adding the Gaussian contribution to the fitted non-Gaussian one, the
underestimation in the diagonal elements is always better than $10\%$, with an
average of $5\%$.

In summary, the fitting formula for the cosmic
variance, including Gaussian and non-Gaussian contributions, has an average
accuracy of $10\%$ in the off-diagonal and $5\%$ in the diagonal.
It is valid when both scales are in the range $50 < \ell < 5000$,
corresponding to $2\arcmin < \theta < 5^{\circ}$ in real
space. This is roughly the range used in the latest results from CFHTLS-Wide 
\citep{2008A&A...479....9F}. This range includes
the scales where non-Gaussianity is relevant, i.e., where the
 cosmic shear error budget is both dominated by cosmic
variance and has important contributions from non-linear clustering (see
Sect.~\ref{sec:fisher}). 

\subsection{Impact of the accuracy of the fitting formula on parameter estimations}
\label{sec:fisher}

We study the impact of the fitting formula accuracy on the estimation of
cosmological parameters, using a Fisher matrix approach. For this, we need to
take into account not only the Gaussian and non-Gaussian contributions to the
covariance, but also the noise in the observed power spectrum. In practical
applications, the convergence field in Eq.~(\ref{eq:estimator}) is obtained
from the observed ellipticities of the source galaxies. The intrinsic
ellipticity field (i.e., in the absence of a gravitational lensing effect) is
assumed to have zero mean and rms of $\sigma_\epsilon$ per component. This
shape noise contaminates the observed power spectrum. Assuming that the
intrinsic ellipticities of different galaxies do not correlate, the shape
noise contribution to the covariance of the power spectrum is diagonal and
given by the new terms arising in Eq.~(\ref{eq:wlcov}) when replacing the
power spectrum, in that expression, by the observed one defined as
\citep{1992ApJ...388..272K},
\begin{equation}
\label{eq:shotnoise}
P^{\rm obs}_\kappa(\ell)=P_\kappa(\ell)+\frac{\sigma^2_\epsilon}{n},
\end{equation}
where $n$ is the number density of source galaxies.

We consider the following three surveys: a medium-deep weak lensing survey
covering an area of $170\deg^2$ with $n=10~{\rm arcmin^2}$, like the current
CFHTLS-Wide, a wider survey of similar depth covering $5000\deg^2$ with
$n=10~{\rm arcmin^2}$, like the planned Dark Energy Survey
(DES)\footnote{\url{http://www.darkenergysurvey.org/}}, and a wide and deep
survey with $20\,000\deg^2$ and $n=40~{\rm arcmin^2}$, like the proposed {\sc
  Euclid}\footnote{\url{http://www.dune-mission.net/}}. For all three we
assume $\sigma_\epsilon=0.3$ and compute the covariance using the halo model
and the developed fit formula in the range $50<\ell<5000$. Additionally, we
add the covariance of a Gaussian contribution using a logarithmic binning with
61 bins in a range from $\ell_{\rm low}=1$ to $\ell_{\rm up}=10^5$. The wider
surveys will measure correlations on scales larger than $\ell=50$, which we do
not consider here for comparison purposes. Note also that for very large
scales the flat-sky approximation breaks down.

\begin{figure}
  \centering
  \includegraphics[angle=0,width=0.49\textwidth]{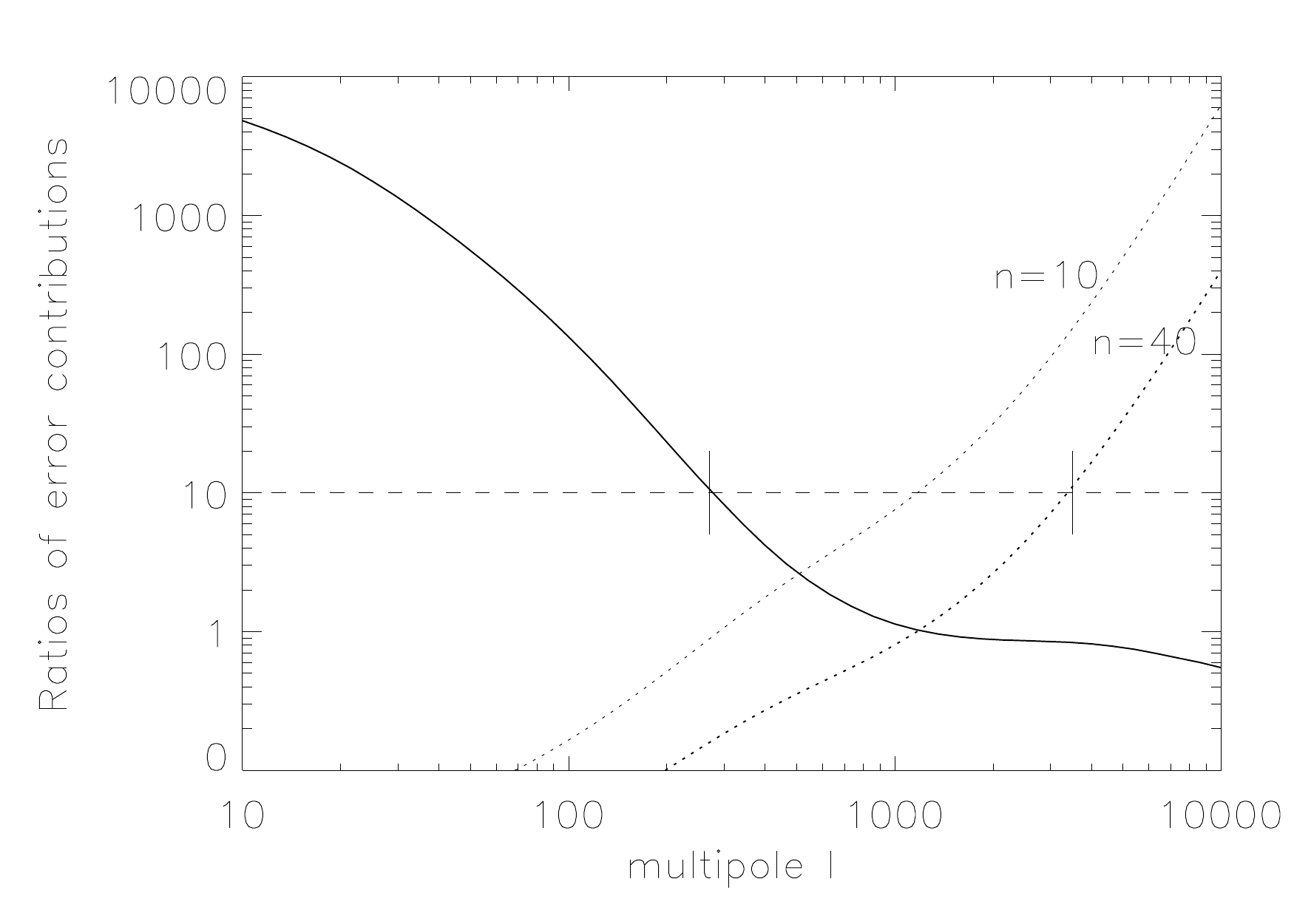}
  \includegraphics[angle=0,width=0.49\textwidth]{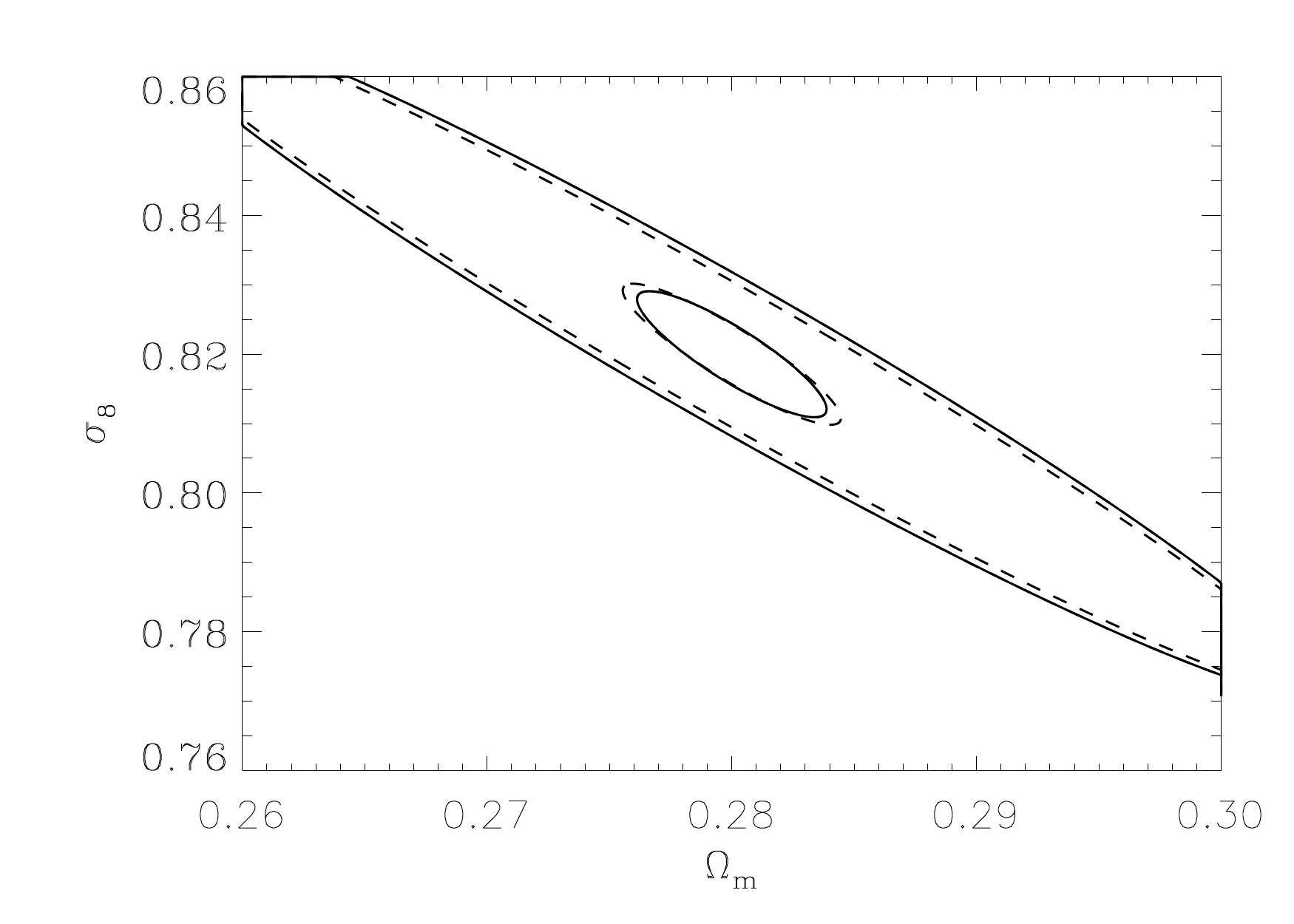}
  \caption{Left panel: Relative contributions to the diagonal of the
    convergence power spectrum covariance. The binning-dependent Gaussian to
    non-Gaussian ratio (solid line; see caption of Fig.~\ref{fig:ratio_fid}
    for the employed binning scheme) and the shape noise to cosmic variance
    ratio for two surveys with $n=10 \,{\rm arcmin}^2$ and $n=40 \,{\rm
      arcmin}^{-2}$(dotted), are shown as function of multipole $\ell$. The
    ticks at the intersection of the $10\%$ line with the ratio curves
    enclose the approximate range where the non-Gaussian contribution is
    important for the {\sc Euclid}-like survey. Right panel: Fisher ellipses
    in the $(\Omega_{\rm m},\sigma_8)$ plane ($2\sigma$ contours) calculated
    with the halo model covariance (solid) and the fitted covariance (dashed)
    for the DES-like (large ellipses) and {\sc Euclid}-like (small ellipses)
    surveys. Compared to the halo covariance case, the fitted covariance error
    ellipses are decreased (enlarged) by $13\%$ ($10\%$) for the DES ({\sc
      Euclid}) case. }
  \label{fig:errbudget}
\end{figure}

The left panel of Fig.~\ref{fig:errbudget} compares the different terms
contributing to the diagonal of the covariance, by showing the Gaussian to
non-Gaussian ratio and the shape noise to cosmic variance ratio, where by
cosmic variance we denote the sum of the Gaussian and non-Gaussian
contributions. The ratio between the Gaussian and non-Gaussian terms is
independent of the survey and, for our particular choice of binning,
non-Gaussianity starts to affect the diagonal around $\ell = 300$, where its
amplitude is $10\%$ of the Gaussian amplitude, and dominates from $\ell
\approx 1000$ onwards. Shape noise, including both pure shape noise and the
coupling with cosmic variance, also becomes important on small scales, but in
a survey-dependent way. It is as large as the cosmic variance on $\ell \approx
200$ $(\ell \approx 1000)$ for surveys with 10 (40) galaxies per arcmin
squared. The vertical lines in Fig.~\ref{fig:errbudget} (left panel) show, for
the {\sc Euclid}-like survey, the range where the non-Gaussian contribution to
the diagonal is non-negligible, i.e., where it accounts for more than $10\%$
of the cosmic variance while having an amplitude of at least $10\%$ of the
shape noise. This range is roughly $300 < \ell < 3000$, or approximately $4
\arcmin < \theta < 50 \arcmin$. This is a rough estimate of the minimum range
where the fitting formula is required to have a good accuracy.

In addition, the accuracy of the off-diagonal terms is
crucial, since the non-Gaussianity is the sole contribution there.
To evaluate the required range of validity of the fitting formula, in a way that includes the
off-diagonal elements and is independent of bin width, we 
define the signal-to-noise ratio \citep{2009MNRAS.395.2065T},
\begin{equation}
\label{eq:snr}
\left(\frac{S}{N}\right)^2=\sum_{ij}\mathcal{P}_{\kappa}(\ell_i)
 \,{\mathcal{C}^{-1}_{ij}}\,\mathcal{P}_{\kappa}(\ell_j)\,.
\end{equation}
For each survey, we compute the signal-to-noise ratio (SNR) using all scales
between $\ell=50$, the largest scale where the fitting is valid, to successive
values of $\ell_{\rm max}$. The SNR increases with $\ell_{\rm max}$, as more
scales are included in Eq.~(\ref{eq:snr}). The increasing rate, however,
decreases with increasing $\ell_{\rm max}$, tending to zero when additional
scales do not carry additional cosmological information. The value of
$\ell_{\rm max}$ for which this saturation occurs increases with decreasing
shape noise, and gives a good indication of the range where the fitting
formula is required to have a good accuracy in order to get accurate estimates
of cosmological parameters. We find that, when increasing $\ell_{\rm max}$
from $\ell_{\rm{max}}=5000$ to $\ell_{\rm{max}}=10000$, the SNR increases by a
factor of 1.03 for the CFHTLS-like survey and by a factor of 1.09 for the {\sc
  Euclid}-like survey. Hence, even for the latter there is no much gain in
reaching $\ell > 5000$.
 
We consider now the Fisher information matrix, which to first order,
neglecting the cosmology dependence of the covariance matrix, is given by
\begin{equation}
  \label{eq:fisher}
  F_{\alpha\beta}=\sum_{ij}\frac{\partial \mathcal{P}_\kappa(\ell_i)}{\partial
    p_\alpha}  \,{\mathcal{C}^{-1}_{ij}}\,\frac{\partial \mathcal{P}_\kappa(\ell_j)}{\partial p_\beta}\,,
\end{equation}
where the derivatives are taken w.r.t. a set of cosmological parameters
$p_\alpha$. The Fisher matrix defines the error ellipsoid in parameter
space, with $\left(F^{-1}_{\,\alpha\alpha}\right)^{1/2}$ yielding the $1\sigma$ marginalized error
on $p_\alpha$, and $\left(1/F_{\alpha\alpha}\right)^{1/2}$ giving the $1\sigma$ error
on $p_\alpha$ assuming the other parameters are perfectly known.

We compute Eq.~(\ref{eq:fisher}) for both the fit and halo model
covariances, for each of the three surveys. We perform the
derivatives at the fiducial model of Tab.~\ref{tab:fiducial_model}, varying  
only 2 cosmological parameters $(\Omega_{\rm m},\sigma_8)$.
For each survey, we compare the areas of the two $2\sigma$ error ellipses
(which defines the inverse of the figure-of-merit) thus obtained. For all
three cases, there is a good agreement between the two ellipses.

For the two cases with large shape noise, CFHTLS and DES, we find that the
fitting formula underestimates the Fisher ellipse, as compared to the halo
model covariance. This is expected because with an enhanced diagonal the
correlations between bins are weaker, and the result is dominated by the
accuracy of the diagonal, where the fitting formula underestimates the
covariance, as we saw earlier on. The deviation is however weak, the areas of
the ellipses obtained using the fitting formula are $13\%$ smaller than the
halo model result, and the deviation is uniformly distributed on the parameter
space (see Fig.~\ref{fig:errbudget}, right panel, large ellipses). This
implies that the deviation on the marginalized constraints is much smaller,
and we find that the fitting formula underestimates the errors on both
$\Omega_{\rm m}$ and $\sigma_8$ by only $1\%$, for both surveys. This
corresponds to a deviation of $0.1\%$ $(0.01\%)$ of the parameters values, for
CFHTLS (DES). In contrast, for the {\sc Euclid} case, where the covariance has
larger correlations, the fitted covariance produced an ellipse slightly larger
than the halo model one, by about $10\%$ (see Fig.~\ref{fig:errbudget}, right
panel, small ellipses).

\subsection{Covariance of real-space estimators}
\label{sec:real-space}

For practical purposes it is sometimes more convenient to study
real-space correlations rather than correlations in Fourier space. We
therefore define an estimator of a general second-order cosmic shear
measure which is related to the convergence power spectrum estimator
by
\begin{equation}
  \label{eq:est-gamma}
  \hat{\Gamma}(\theta)=\int_{0}^{\infty}\frac{\abl
    \ell\,\ell}{2\pi}W(\ell\theta)\hat{P}_{\kappa}(\ell)\,,
\end{equation}
where $W(x)$ is an arbitrary weight function. A well-known example of this
equation are the shear two-point correlation functions $\xi_{+}(\theta)$ and
$\xi_{-}(\theta)$ with weight functions
$W(\ell\theta)=\operatorname{J}_{0}(\ell\theta)$ and
$W(\ell\theta)=\operatorname{J}_{4}(\ell\theta)$, respectively (see discussion
in \citealt{2008A&A...477...43J}). Using the definition
Eq.~\eqref{eq:est-gamma}, we find for the relation between the covariance of
the real-space estimators to the covariance of the Fourier-space estimators:
\begin{equation}
  \operatorname{Cov}\left[\hat{\Gamma}(\theta),\hat{\Gamma}(\theta')\right]=
  \int_{0}^{\infty}\frac{\abl \ell\,\ell}{2\pi}W(\ell\theta) 
  \int_{0}^{\infty}\frac{\abl \ell'\,\ell'}{2\pi}W(\ell'\theta')
  \operatorname{Cov}\left[\hat{P}_{\kappa}(\ell),\hat{P}_{\kappa}(\ell')\right]\,,
\end{equation}
which is related to the covariance of the dimensionless power spectrum used in
the fitting formula by
$(2\pi)^{2}\operatorname{Cov}\left[\hat{\mathcal{P}}_{\kappa}(\ell),\hat{\mathcal{P}}_{\kappa}(\ell')\right]=\ell^{2}\ell'^{2}
\operatorname{Cov}\left[\hat{P}_{\kappa}(\ell),\hat{P}_{\kappa}(\ell')\right]$.
Inserting the result of the dimensionless power spectrum covariance given in
Eq.~\eqref{eq:wlcov} yields
\begin{equation}
  \operatorname{Cov}\left[\hat{\Gamma}(\theta),\hat{\Gamma}(\theta')\right]=
  \frac{4\pi}{A}\int_{0}^{\infty}\frac{\abl\ell}{\ell^{3}}\,\mathcal{P}_{\kappa}^{2}(\ell)
  W(\ell\theta)W(\ell\theta')
  +
  \frac{1}{A}\int_{0}^{\infty}\frac{\abl\ell}{\ell}\,W(\ell\theta)
  \int_{0}^{\infty}\frac{\abl\ell'}{\ell'}\,\bar{T}_{\kappa}(\ell,\ell') W(\ell'\theta')\,.
\end{equation}
We find that the Gaussian part of the real-space covariance is
independent of the binning scheme and is non-diagonal in contrast to
the covariance in Fourier space.

\section{Conclusions}
\label{sec:discussion}

We present a fitting formula for the halo model prediction of the non-Gaussian
contribution to the covariance of the dimensionless power spectrum of the weak
lensing convergence. The formula was constructed assuming a $\Lambda$CDM
cosmology with WMAP5-like cosmological parameters. In particular, it was
obtained for $\Omega_{\rm m}=0.28$, $\sigma_8=0.82$ and other parameter values
as shown in Tab.~\ref{tab:fiducial_model}. It is valid for a scale range of
$50 \lesssim \ell \lesssim 5000$, corresponding to $2'<\theta<5^{\circ}$ in
real space and can be used for surveys with galaxy source redshifts $z_{\rm s}
\in [0.5,2]$. In this range, it reproduces the results of a full
implementation of the halo model approach, with a scale-averaged accuracy of
$10\%$ in the off-diagonal, and $5\%$ in the diagonal elements. The formula
also allows us to recover the halo model $(\Omega_{\rm m},\sigma_8)$ error
ellipses within $15\%$. The range of validity of the formula and its level of
accuracy render it applicable to low shape noise scenarios from next
generation weak lensing surveys.

To use the formula, shown in Eq.~(\ref{eq:fitting_formula}), one needs
three quantities~:
\begin{itemize}
\item The non-Gaussian contribution to the covariance of the
  convergence power spectrum in tree-level perturbation theory ${\cal
    C}_{\rm pt}$, shown in Eq.~(\ref{eq:cov_perturbation}). This
  involves the computation of the convergence trispectrum in
  tree-level perturbation theory, Eq.~(\ref{eq:tricov}), which
  requires the calculation of the linear power spectrum and of the
  $F_2$ and $F_3$ coupling functions (Eqs.~\ref{eq:F2} and
  \ref{eq:F3}).
\item
The non-linear convergence power spectrum.
\item
The 9 coefficients of the fit, which are obtained by inserting the 27 values
given in Tab.~\ref{tab:fitaz} in Eq.~(\ref{eq:fitaz}), for the required redshift.
\end{itemize}
The Gaussian contribution, calculated from the non-linear convergence
power spectrum as given in Eq.~(\ref{eq:wlcov}), may then be added to
the result of the formula, to obtain the total covariance. This is the
covariance of the estimator without noise, or the cosmic variance.
 
The work presented in this paper is based on the assumption that the
halo model is a powerful approach to probe non-linear clustering. We
tried to test this assumption against results from $N$-body
simulations, but our comparisons were inconclusive. Indeed, such
analysis requires ray-tracing simulations with both large number of
convergence maps and large convergence map area. Only then it would be
possible to minimize the effect of sampling variance in the
simulations, which hindered our attempted tests. Such simulations
would also allow us to consider error bars for the estimate of the
simulation covariance. There are however indications that the halo model
approach underestimates the non-Gaussianity of the covariance and there are 
attempts to include additional contributions \citep{2009MNRAS.395.2065T,2009arXiv0906.2237S}.

The reliability of the halo model for higher-order polyspectra also needs to
be studied in more detail. Some work in this direction are the analyses of the
impact of the triaxiality of the halo profiles
\citep{2005MNRAS.360..203S,2006MNRAS.365..214S}, and halo exclusion effects
\citep{2005ApJ...631...41T}. Moreover, the issue of halo substructure
(\citealt{2004MNRAS.352.1019D}) and of the effect of a stochastic
concentration parameter have to be understood properly. This paper also
addresses this last issue. We analyze the impact of a stochastic concentration
parameter on the covariance of the convergence power spectrum. We found that
the effect can safely be neglected for the Gaussian contribution, with the
convergence power spectrum varying only slightly, and at small scales, for
concentration scatters of $\sigma_{\ln c} \simeq 0.2-0.3$. For the
non-Gaussian contribution the effect is more pronounced due to the higher
sensitivity of the trispectrum to a stochastic concentration relation. In the
case of the 1-halo term of the trispectrum, we find it useful to take into
account a concentration dispersion of $\sigma_{\ln c} \gtrsim 0.3$. The
deviation to a deterministic concentration-mass relation is larger than $12\%$
for wave-numbers $\ell \gtrsim 3000$.

Although the fitting formula we obtained provides a more thorough
estimate for the non-Gaussian contribution to the power spectrum
covariance than the earlier approximation of
\citet{2007MNRAS.375L...6S} (global accuracy of $\sim 20 \%$ along the
diagonal, which becomes less for the off-diagonals), there is still
room for improvements. One drawback of our approach is that it
requires the computation of the convergence trispectrum in tree-level
perturbation theory. Furthermore, it is only applicable to a small
range of WMAP5-like cosmologies and is only valid in the interval $50
\leq \ell\leq 5000$. A possible way to avoid these problems and extend
the accuracy of the fitting formula might be to construct it entirely
from its three-dimensional counterpart, the three-dimensional
covariance of the matter power spectrum. This has the advantage that
perturbations of different length scales are not additionally mixed
due to projection effects, which might allow us to cover a wider range
of cosmologies. The desired projected covariance could then be
obtained by performing an additional integration along the
redshift-space. We will address this issue in a future paper.

\begin{acknowledgements}
  The authors thank Christoph Lampert for invaluable discussions, and
  Jan Hartlap for providing his ray-tracing simulations of the Gems
  and Virgo simulations. Together with Martin Kilbinger both provided
  useful comments to the manuscript. JP would like to thank the TRR33.
  JR is supported by the Deutsche Forschungsgemeinschaft under the
  project SCHN 342/7--1 within the Priority Programme SPP 1177 `Galaxy
  Evolution'. IT is supported by the Marie Curie Training and Research
  Network `DUEL'.
\end{acknowledgements}

\appendix
\section{Cosmological Perturbation Theory}
\label{sec:PT}

On large scales, different Fourier modes evolve independently from
each other and thus conserve the Gaussian behavior of the density
perturbation field $\tilde{\delta}(\bs k,a)$. It is therefore
convenient to work in Fourier space and Fourier transform the fields
as well as the non-linear fluid equations (consisting of continuity,
Euler and Poisson equation) that describe their evolution in an
expanding Universe. Contrary to linear perturbation theory, there is a
coupling between different Fourier modes mediated by the coupling
function $\alpha(\bs k_{1},\bs k_{2})$ on smaller scales. In this
case, the fluid equations for the density contrast and the irrotational
peculiar velocity field $\theta=\nabla\cdot \bs u$ in Fourier space are given by
\citep[e.g.,][]{LSS_PT}
\begin{align}
\label{eq:ContFourier}
a \dot{\tilde{\delta}}(\bs k, a) + \tilde{\theta}(\bs k,a) &= - \int
\frac{\md^3 k_1}{(2\pi)^3} \int \md^3 k_2 \delta_{\rm D}(\bs k - \bs
k_1 - \bs k_2) \alpha(\bs k_1, \bs k_2) \tilde{\theta}(\bs k_1,a)
\tilde{\delta}(\bs k_2,a) \,,  \\
\label{eq:EulerFourier}
a \dot{\tilde{\theta}}(\bs k,a) + \frac{3H_0^2 \Omega_{\rm m}}{2a}
\, \tilde{\delta}(\bs k,a) &= - \int \frac{\md^3 k_1}{(2\pi)^3} \int \md^3
k_2 \delta_{\rm D}(\bs k - \bs k_1 - \bs k_2) \beta(\bs k_1,\bs k_2)
\tilde{\theta}(\bs k_1,a) \tilde{\theta}(\bs k_2,a) \,,
\end{align}
where we introduced the two fundamental mode coupling functions
\begin{align}
  \label{eq:alpha}
  \alpha(\bs k_{1},\bs k_{2})=
  \frac{(\bs k_{1}+\bs k_{2})\cdot \bs k_{1}}{k_{1}^{2}} \,,\qquad
  \beta(\bs{k_1},\bs{k_2}) =
  \frac{|\bs{k_1}+\bs{k_2}|^2(\bs{k_1}\cdot\bs{k_2})}{2k_1^2 k_2^2}\,.
\end{align}
For an Einstein-de Sitter (EdS) cosmology it is possible to find a
perturbative ansatz that separates the scale- and time dependencies, whereas
for a general $\Lambda$CDM model it is impossible to find a separable solution
to Eqs.~\eqref{eq:ContFourier} and \eqref{eq:EulerFourier}. However,
\citet{1998ApJ...496..586S} showed that it is possible to find a separable
solution in any order if one makes an approximation that is valid at
percentage level. One indeed finds then the same recursion relation as in the
EdS case. The ansatz is then
\begin{equation}
  \label{eq:PT-ansatz}
  \tilde{\delta}(\bs k,a) = \sum_{n=1}^\infty
  D^n(a)\tilde{\delta}_n(\bs k) \,, \qquad \tilde{\theta}(\bs k,a) = -
  \dot{a} \sum_{n=1}^\infty D^n(a) \tilde{\theta}_n(\bs k) \,.
\end{equation}
Therefore, we find that the whole information on cosmological
parameters is encoded in the growth function due to its dependence on
the Hubble parameter (see Eq.~\ref{eq:growth_factor}).

\subsection{Coupling Functions}
\label{sec:coupling-functions}

The $n$-th order density contrast and the divergence of the peculiar velocity
in Eq.~\eqref{eq:PT-ansatz} is given by
\begin{align}
  \label{eq:defdelta_n}
  \tilde\delta_{n}(\bs k)&=\int \frac{\abl^{3}q_{1}}{(2\pi)^{3}}\cdots
  \frac{\abl^{3}q_{n-1}}{(2\pi)^{3}}\int\abl^{3}q_{n} \,\delta_{\rm
    D}(\bs k -\bs q_{1\ldots n })
  F_{n}(\bs q_{1},\ldots,\bs q_{n}) \tilde\delta_{1}(\bs q_{1})\cdots \tilde\delta_{1}(\bs q_{n})  \,,\\
  \tilde\theta_{n}(\bs k)&=\int \frac{\abl^{3}q_{1}}{(2\pi)^{3}}\cdots
  \frac{\abl^{3}q_{n-1}}{(2\pi)^{3}} \int\abl^{3}q_{n} \, \delta_{\rm
    D}(\bs k -\bs q_{1\ldots n}) G_{n}(\bs q_{1},\ldots,\bs q_{n})
  \tilde\delta_{1}(\bs q_{1})\cdots \tilde\delta_{1}(\bs q_{n}) \,,
\end{align}
and the $n$-th order coupling functions $F_n$ and $G_n$ are obtained by the
following recursion relations (\citealt{1994ApJ...431..495J})
\begin{align}
F_n(\bs{q_1},\ldots,\bs{q_n})&=\sum_{m=1}^{n-1} 
\frac{G_m(\bs{q_1},\ldots,\bs{q_m})}{(2n+3)(n-1)}
[(2n+1)\alpha(\bs{k_1,\bs{k_2}}) F_{n-m}(\bs{q_{m+1}},\ldots,\bs{q_n})
 + 2\beta(\bs{k_1},\bs{k_2})G_{n-m}(\bs{q_{m+1}},\ldots,\bs{q_n})]\,,\\
G_n(\bs{q_1},\ldots,\bs{q_n})&=\sum_{m=1}^{n-1}
\frac{G_m(\bs{q_1},\ldots,\bs{q_m})}{(2n+3)(n-1)}
[3\alpha(\bs{k_1,\bs{k_2}}) F_{n-m}(\bs{q_{m+1}},\ldots,\bs{q_n})
+ 2n\beta(\bs{k_1},\bs{k_2})G_{n-m}(\bs{q_{m+1}},\ldots,\bs{q_n})]\,,
\end{align}
where $\bs{k_1}\equiv \bs{q_1}+\ldots+\bs{q_m}$ and $\bs{k_2}\equiv
\bs{q_{m+1}}+\ldots+\bs{q_n}$. The initial conditions for these recursion
relations are $F_1\equiv 1$ and $G_1\equiv 1$. 
To get the functions $F_n^{(\textrm{s})}$ and $G_n^{(\textrm{s})}$ that are
symmetric in its arguments, one must perform the following symmetrizing
procedure
\begin{align}
\label{eq:symtrize}
F_n^{(\textrm{s})}(\bs{q_1},\ldots,\bs{q_n})= \frac{1}{n!}\sum_\pi
F_n(\bs{q_{\pi(1)}},\ldots,\bs{q_{\pi(n)}})\,,\qquad
G_n^{(\textrm{s})}(\bs{q_1},\ldots,\bs{q_n})= \frac{1}{n!}\sum_\pi
G_n(\bs{q_{\pi(1)}},\ldots,\bs{q_{\pi(n)}})\,,
\end{align}
where the sum is taken over all possible permutations $\pi$ of the set $\{
1,\ldots,n\}$. These equations enable us to calculate the density contrast in
the $n$-th order of perturbation theory by using the iterative equations for
the coupling functions. 

The calculation of the second-order coupling functions is straightforward. The
result is
\begin{align}
\label{eq:F2}
F_2^{(\textrm{s})}(\bs{q_1},\bs{q_2}) = \frac{5}{7}+
\frac{2}{7}\frac{(\bs{q_1}\cdot\bs{q_2})^2}{q_1^2 q_2^2}
+ \frac{1}{2}\frac{\bs{q_1}\cdot\bs{q_2}}{q_1 q_2}
\left(\frac{q_1}{q_2}+\frac{q_2}{q_1}\right)\,,\qquad
G_2^{(\textrm{s})}(\bs{q_1},\bs{q_2}) = \frac{3}{7}+
\frac{4}{7}\frac{(\bs{q_1}\cdot\bs{q_2})^2}{q_1^2 q_2^2}
+ \frac{1}{2}\frac{\bs{q_1}\cdot\bs{q_2}}{q_1 q_2}
\left(\frac{q_1}{q_2}+\frac{q_2}{q_1}\right) \, .
\end{align}
The third-order coupling function is given by
\begin{align}
F_3(\bs{q_1},\bs{q_2},\bs{q_3}) = \frac{1}{18}\Big\{
7\alpha(\bs{q_1},\bs q_{23})F_2(\bs{q_2},\bs{q_3})+ 
2\beta(\bs{q_1},\bs q_{23})G_2(\bs{q_2},\bs{q_3}) 
 + [7\alpha(\bs q_{12},\bs{q_3}) +
2\beta(\bs q_{12},\bs{q_3})] G_2(\bs{q_1},\bs{q_2})\Big\} \,,
\end{align}
where $\bs q_{ij}\equiv \bs q_{i}+\bs q_{j}$. Employing
Eq.~\eqref{eq:symtrize}, we find the symmetric function
\begin{align}
  F_3^{(\textrm{s})}(\bs{q_1},\bs{q_2},\bs{q_3}) &=
\frac{7}{54}
[\alpha(\bs{q_1},\bs q_{23})F_2^{(\textrm{s})}(\bs{q_2},\bs{q_3})+
\alpha(\bs{q_2},\bs q_{13})F_2^{(\textrm{s})}(\bs{q_1},\bs{q_3}) 
  +\alpha(\bs{q_3},\bs q_{12})F_2^{(\textrm{s})}(\bs{q_1},\bs{q_2})]
\notag\\
  &{}\quad+ \frac{4}{54}
[\beta(\bs{q_1},\bs q_{23})G_2^{(\textrm{s})}(\bs{q_2},\bs{q_3})+
\beta(\bs{q_2},\bs q_{13})G_2^{(\textrm{s})}(\bs{q_1},\bs{q_3}) 
 +\beta(\bs{q_3},\bs q_{12})G_2^{(\textrm{s})}(\bs{q_1},\bs{q_2})]
\notag\\
  &{}\quad+\frac{7}{54}
[\alpha(\bs q_{12},\bs{q_3})G_2^{(\textrm{s})}(\bs{q_1},\bs{q_2})+
\alpha(\bs q_{13},\bs{q_2})G_2^{(\textrm{s})}(\bs{q_1},\bs{q_3}) 
 +\alpha(\bs q_{23},\bs{q_1})G_2^{(\textrm{s})}(\bs{q_2},\bs{q_3})] \,.
\label{eq:F3}
\end{align}
From now on the symmetry superscript ``$(\textrm{s})$'' will be omitted
because we will only deal with symmetric coupling functions. For the
calculation of the trispectrum in the halo model approach as described
in Sect.~\ref{sec:trispectrum-2}, one needs perturbation theory. More
precisely, we need the subsequent components
\begin{align}
F_3(\bs{q_1},-\bs{q_1},\bs{q_2}) &= \frac{7}{54}
[\alpha(\bs{q_1},\bs q_{-})F_2(-\bs{q_1},\bs{q_2})+
\alpha(-\bs{q_1},\bs q_{+})F_2(\bs{q_1},\bs{q_2})] \notag\\
&{}\quad + \frac{4}{54}
[\beta(\bs{q_1},\bs q_{-})G_2(-\bs{q_1},\bs{q_2})+
\beta(-\bs{q_1},\bs q_{+})G_2(\bs{q_1},\bs{q_2})] \notag\\
&{}\quad +\frac{7}{54}
[\alpha(\bs q_{-},\bs{q_1})G_2(-\bs{q_1},\bs{q_2})+
\alpha(\bs q_{+},-\bs{q_1})G_2(\bs{q_1},\bs{q_2})] \,, 
\end{align}
where we have defined the difference vector $\bs q_{-}\equiv \bs q_{2}-\bs
q_{1}$ and the sum of the vectors $\bs q_{+}\equiv \bs q_{1}+\bs q_{2}$.  

We already mentioned in the previous section that it is possible to find a
solution for an arbitrary cosmology if one makes a small approximation. In the
literature one can find closed solutions for the second- and third-order coupling
functions. The second-order coupling function changes to
\begin{equation}
  F_{2}(\bs q_{1},\bs q_{2})=\frac{1}{2}(1+\epsilon)+\frac{1}{2}\frac{\bs
    q_{1}\cdot \bs q_{2}}{q_{1}q_{2}}\left(\frac{q_{1}}{q_{2}}+
    \frac{q_{2}}{q_{1}}\right)+
  \left(\frac{1}{2}-\frac{\epsilon}{2}\right)\frac{(\bs q_{1}\cdot\bs
    q_{2})^{2}}{q_{1}^{2}q_{2}^{2}} \,,
\end{equation}
where $\epsilon\approx (3/7)\Omega_{\rm m}^{-2/63}$ for $\Omega_{\rm m}\gtrsim
0.1$ (\citealt{LSS_PT}). For our fiducial choice of $\Omega_{\rm
  m}=0.3$, we get $\Omega_{\rm m}^{-2/63}\approx 1.039$. Thus, within a few
percent correction to the first and last term, the coupling functions are
independent of cosmological parameters.

\subsection{Correlation functions}
\label{sec:correlation-function}

\subsubsection{Bispectrum}

The dark matter bispectrum is defined as
\begin{equation}
  \label{eq:defbispectrum}
  \langle \tilde\delta(\bs k_{1}) \tilde\delta(\bs
  k_{2})\tilde\delta(\bs k_{3})\rangle_{\rm c} = (2\pi)^{3}\delta_{\rm
    D}(\bs k_{123}) B(\bs k_{1},\bs k_{2},\bs k_{3})\,.
\end{equation}
Since the connected bispectrum vanishes for Gaussian random fields, it is the
first intrinsically non-linear moment. Inserting the perturbative expansion
\eqref{eq:PT-ansatz} for each term results generally in an infinitely large
sequence of correlators. The lowest non-vanishing order is the so-called
\emph{tree-level} contribution to the bispectrum. We find for the correlator
in tree level
\begin{align}
  \langle \tilde\delta(\bs k_{1})\tilde\delta(\bs k_{2})
  \tilde\delta(\bs k_{3})\rangle_{\rm tree} =
  \langle\tilde\delta_{2}(\bs k_{1})\tilde\delta_{1}(\bs k_{2})
  \tilde\delta_{1}(\bs k_{3})\rangle +
  \langle\tilde\delta_{1}(\bs k_{1})\tilde\delta_{2}(\bs k_{2})
  \tilde\delta_{1}(\bs k_{3})\rangle 
   + 
  \langle\tilde\delta_{1}(\bs k_{1})\tilde\delta_{1}(\bs k_{2})
  \tilde\delta_{2}(\bs k_{3})\rangle  \,.
\end{align}
Replacing the second-order density contrast with Eq.~\eqref{eq:defdelta_n}
results in
\begin{align}
  \langle\tilde\delta_{2}(\bs k_{1})\tilde\delta_{1}(\bs
  k_{2})\tilde\delta_{1}(\bs k_{3})\rangle
  &=\int\frac{\abl^{3}q_{1}}{(2\pi)^3}\int\abl^{3}q_{2}\,
  \delta_{\rm{D}}(\bs k_{1}-\bs q_{1}-\bs q_{2})F_{2}(\bs q_{1},\bs
  q_{2}) \langle\tilde\delta_{1}(\bs q_{1})\tilde\delta_{1}(\bs
  q_{2})\tilde\delta_{1}(\bs
  k_{2})\tilde\delta_{1}(\bs k_{3})\rangle\notag\\
  &=(2\pi)^{3}\delta_{\rm D}(\bs k_{123}) \, 2F_{2}(\bs k_{2},\bs
  k_{3})P_{\rm{lin}}(k_{2})P_{\rm{lin}}(k_{3})\,,
\end{align}
where we applied Wick's theorem to express the four-point correlator of
Gaussian fields in terms of products of power spectra, and performed the two
integrations over the Dirac delta distributions. The results for the other two
terms of the tree-level bispectrum are simply obtained by permutations of the
arguments. Finally, the tree-level bispectrum is given by
\begin{align}
  \label{eq:linbispectrum}
  B_{\rm{pt}}(\bs k_{1},\bs k_{2},\bs k_{3})=
  2F_{2}(\bs k_{1},\bs k_{2})P_{1}P_{2}+ 
  2F_{2}(\bs k_{1},\bs k_{3})P_{1}P_{3} 
  + 2F_{2}(\bs k_{2},\bs k_{3})P_{2}P_{3} \,.
\end{align}
The factor 2 follows from using the symmetrized version of the second-order
coupling function.

\subsubsection{Trispectrum}

The dark matter trispectrum is defined as the connected four-point function in
Fourier space:
\begin{equation}
  \label{eq:deftrispectrum}
  \langle \tilde\delta(\bs k_{1}) \tilde\delta(\bs k_{2})
  \tilde\delta(\bs k_{3})\tilde\delta(\bs k_{4}) \rangle_{\rm c} =
  (2\pi)^{3}\delta_{\rm D}(\bs k_{1234})
  T(\bs k_{1},\bs k_{2},\bs k_{3},\bs k_{4}) \,,
\end{equation}
where $\bs k_{1234}\equiv \bs k_{1}+\bs k_{2}+\bs k_{3}+\bs k_{4}$. We find
that there are two different non-vanishing contributions to the tree level:
\begin{align}
  \langle \tilde\delta(\bs k_{1}) 
  \tilde\delta(\bs k_{2})\tilde\delta(\bs k_{3})\tilde\delta(\bs k_{4})
  \rangle_{\rm tree} &= 
  \langle \tilde\delta_{2}(\bs k_{1}) \tilde\delta_{2}(\bs k_{2}) 
  \tilde\delta_{1}(\bs k_{3})\tilde\delta_{1}(\bs k_{4})
  \rangle + \ldots (\textrm{6 terms}) \notag\\
  &{}\quad +\langle \tilde\delta_{3}(\bs k_{1}) 
  \tilde\delta_{1}(\bs k_{2})\tilde\delta_{1}(\bs k_{3})
  \tilde\delta_{1}(\bs k_{4}) \rangle +\ldots (\textrm{4 terms}) \,.
\end{align}
In total we find 6 terms of the first type and 4 terms for the second type,
where the rest is obtained by permutations. All other contributions either
vanish or are built of higher-order terms. Note that for the second type of
terms we need the results from perturbation theory up to the third order.
The calculation of each term is a tedious but straightforward calculation. We
obtain for the first term of the expansion
\begin{align}
  \langle \tilde\delta_{2}(\bs k_{1}) \tilde\delta_{2}(\bs k_{2})
  \tilde\delta_{1}(\bs k_{3})\tilde\delta_{1}(\bs
  k_{4})\rangle=(2\pi)^{3}\delta_{\rm D}(\bs k_{1234}) \,
  4P_{3}P_{4}[P_{13}F_{2}(\bs k_{3},-\bs k_{13})F_{2}(\bs k_{4},-\bs
  k_{24})+ P_{14}F_{2}(\bs k_{3},-\bs k_{23})F_{2}(\bs k_{4},-\bs
  k_{14})]\,,
\end{align}
where the six-point correlator resolves into 15 terms consisting of power
spectra products. Performing the integrations over the arising delta functions
yields in the end 8 different terms. Similarly, we find for the second type of
terms
\begin{align}
  \langle \tilde\delta_{3}(\bs k_{1}) \tilde\delta_{1}(\bs
  k_{2})\tilde\delta_{1}(\bs k_{3}) \tilde\delta_{1}(\bs k_{4})
  \rangle= (2\pi)^{3}\delta_{\rm D}(\bs k_{1234}) \, 6F_{3}(\bs
  k_{2},\bs k_{3},\bs k_{4})P_{2}P_{3}P_{4}\,.
\end{align}
The other terms are easily obtained by permutations, however, we present here
the complete result to avoid confusion with a shorthand notation that is
introduced afterwards. The trispectrum of cold dark matter is in first
non-vanishing order given by (\citealt{1984ApJ...279..499F}):
\begin{equation}
  \label{eq:triPT}
  T_{\rm{pt}} = 4 T_a+ 6 T_b \,,
\end{equation}
where 
\begin{align}
  \label{eq:Tafull}
  T_a &= 
  P_1 P_2 \left[ 
    P_{13} F_2(\bs k_1,-\bs k_{13}) F_2(\bs k_2,\bs k_{13}) + 
    P_{14} F_2(\bs k_1,-\bs k_{14}) F_2(\bs k_2,\bs k_{14})  
  \right] \notag\\ 
  &{}\quad+ P_1 P_3 \left[ 
    P_{12} F_2(\bs k_1,-\bs k_{12}) F_2(\bs k_3,\bs k_{12}) + 
    P_{14} F_2(\bs k_1,-\bs k_{14}) F_2(\bs k_3,\bs k_{14})  
  \right]  \notag \\
  &{}\quad+ P_1 P_4 \left[ 
    P_{12} F_2(\bs k_1,-\bs k_{12}) F_2(\bs k_4,\bs k_{12}) + 
    P_{13} F_2(\bs k_1,-\bs k_{13}) F_2(\bs k_4,\bs k_{13})  
  \right]  \notag \\
  &{}\quad+ P_2 P_3 \left[ 
    P_{21} F_2(\bs k_2,-\bs k_{21}) F_2(\bs k_3,\bs k_{21}) + 
    P_{24} F_2(\bs k_2,-\bs k_{24}) F_2(\bs k_3,\bs k_{24})  
  \right]  \notag \\
  &{}\quad+ P_2 P_4 \left[ 
    P_{21} F_2(\bs k_2,-\bs k_{21}) F_2(\bs k_4,\bs k_{21}) + 
    P_{23} F_2(\bs k_2,-\bs k_{23}) F_2(\bs k_4,\bs k_{23})  
  \right]  \notag \\
  &{}\quad+ P_3 P_4 \left[ 
    P_{31} F_2(\bs k_3,-\bs k_{31}) F_2(\bs k_4,\bs k_{31}) + 
    P_{32} F_2(\bs k_3,-\bs k_{32}) F_2(\bs k_4,\bs k_{32})  
  \right] \,, 
\end{align}
and
\begin{align}
\label{eq:tbfull}
  T_b = F_3(\bs k_1,\bs k_2,\bs k_3) P_1 P_2 P_3 + 
  F_3(\bs k_2,\bs k_3,\bs k_4) P_2 P_3 P_4  
  + F_3(\bs k_3,\bs k_4,\bs k_1) P_3 P_4 P_1 +
  F_3(\bs k_4,\bs k_1,\bs k_2) P_4 P_1 P_2 \,,
\end{align}
where $P_{i}\equiv P_{\rm{lin}}(k_{i})$, $P_{ij}\equiv P_{\rm{lin}}(|\bs
k_{i}+\bs k_{j}|)$ and $\bs k_{ij}\equiv \bs k_{i}+\bs k_{j}$. 

For the covariance matrix one only needs the parallelogram configuration. This
imposes the condition $\bs k_{2}=-\bs k_{1}$ and $\bs k_{4}=-\bs k_{3}$ on the
wave-vectors. In this case Eq.~\eqref{eq:triPT} simplifies to
\begin{align}
  T_{\rm{pt}}%(\bs{k_1},-\bs{k_1},\bs{k_3},-\bs{k_3})
  &= 
  4 P_{1}^{2} \left\{ 
    [F_2(\bs{k_1},-\bs k_{+})]^2 P_{+}\right. 
    + \left.
   [F_2(\bs{k_1}, \bs k_{-})]^2  P_{-}
  \right\} +4 P_{3}^{2} \left\{
    [F_2(\bs{k_3},-\bs k_{+})]^2 P_{+} \right.
    + \left. 
   [F_2(\bs{k_3},- \bs k_{-})]^2 P_{-} 
    \right\} \notag\\
  &{}\quad+ 8 P_{1} P_{3} 
  \left[F_2(\bs{k_1},-\bs k_{+}) 
  F_2(\bs{k_3},-\bs k_{+}) P_{+} + F_2(\bs{k_1},\bs k_{-}) 
  F_2(\bs{k_3},-\bs k_{-}) P_{-}\right] \notag\\
  &{}\quad+ 12 \left[P_1^2 P_3 
    F_3(\bs{k_1},-\bs{k_1},\bs{k_3}) 
  + P_1 P_3^{2} F_3(\bs{k_1},\bs{k_3},-\bs{k_3})\right] \,,
  \label{eq:tricov}
\end{align}
where  $\bs k_{-}\equiv\bs
k_{3}-\bs k_{1}$, $\bs k_{+}\equiv\bs k_{1}+\bs k_{3}$, $P_{-}\equiv
P_{\rm{lin}}(|\bs k_{-}|)$ and $P_{+}\equiv P_{\rm{lin}}(|\bs
k_{+}|)$. Consequently, we define the non-Gaussian contribution to the
covariance in tree-level perturbation theory as
\begin{equation}
  \label{eq:cov_perturbation}
  {\cal C}_{\rm pt} \equiv {\cal C}_{\rm pt}(\ell_{i},\ell_{j}) = \rez{A}
  \int_{{|\bs l_1|} \in \ell_i} \frac{\abl^{2}\ell_{1}}{A_{\rm{r}}(\ell_{i})}
  \int_{{|\bs l_2|} \in \ell_j} \frac{\abl^{2}\ell_{2}}{A_{\rm{r}}(\ell_{j})}
  \, \frac{\ell_1^2 \ell_2^2}{(2\pi)^2} \, \int_0^{w_{\rm H}} \md w \,
  \frac{ G^4(w)}{w^2} \, T_{\rm pt}\left(\frac{\bs
      \ell_1}{w},-\frac{\bs \ell_1}{w},\frac{\bs \ell_2}{w},-\frac{\bs
      \ell_2}{w},w\right) \,,
\end{equation}
where $A$ denotes the survey area, $A_{\rm{r}}(l)$ the integration area and
$G(w)$ the lensing weight function (see Eq.~\ref{eq:weight_function}). More
details on the notation can be found in Sect.~\ref{sec:con-power-spec-cov}.

\bibliographystyle{aa}
\bibliography{bibdoc}

\end{document}